\documentclass[11pt]{article}

\usepackage[a4paper,margin=1in]{geometry}

\usepackage{tocloft}
\usepackage[utf8]{inputenc}
\usepackage[T1]{fontenc}
\usepackage{amsmath,amssymb,mathrsfs,tensor,bbold,upgreek,cases}
\usepackage{amsthm,amsfonts,mathtools,bm,slashed}
\usepackage[hidelinks,linktoc=all]{hyperref}
\usepackage{url}
\usepackage{xcolor}
\usepackage{comment}
\usepackage{float,graphicx,subfig}
\graphicspath{{figs/}}
\usepackage[font=small,labelfont=bf]{caption}
\usepackage{cite}
\usepackage{algorithm}
\usepackage{algorithmicx}
\usepackage{algpseudocode}

\makeatletter
\g@addto@macro\bfseries{\boldmath}
\makeatother

\makeatletter
\def\blfootnote{\gdef\@thefnmark{}\@footnotetext}
\makeatother

\numberwithin{equation}{section}

\theoremstyle{plain}
\newtheorem{theorem}{Theorem}

\newtheorem{corollary}[theorem]{Corollary}

\theoremstyle{definition}
\newtheorem{definition}[theorem]{Definition}
\newtheorem{assumption}[theorem]{Assumption}

\theoremstyle{remark}

\newcommand\mailto[1]{\href{mailto:#1}{\tt #1}}

\def\be{\begin{equation}}
\def\ee{\end{equation}}

\newcommand{\Q}{\mathcal{Q}}
\newcommand{\M}{\mathcal{M}}
\newcommand{\Lie}{\mathcal{L}}
\newcommand{\q}{q}
\newcommand{\p}{p}
\newcommand{\qq}{\bm{q}}
\newcommand{\pp}{\bm{p}}
\newcommand{\Hc}{\mathscr{H}}

\newcommand{\g}{\psi}
\newcommand{\gv}{\bm{\psi}}
\newcommand{\hh}{\phi}
\newcommand{\hhv}{\bm{\phi}}
\newcommand{\lambdav}{\bm{\lambda}}

\newcommand{\group}{\mathcal{G}}
\newcommand{\liealg}{\mathfrak{g}}

\newcommand{\Jac}{\mathcal{J}}

\newcommand{\Ppar}{{\mathcal{P}_{\parallel}}}
\newcommand{\Pper}{{\mathcal{P}_{\perp}}}

\DeclareMathOperator{\Tr}{tr}

\newcommand{\bigO}{\mathcal{O}}

\begin{document}

\renewcommand{\thefootnote}{\fnsymbol{footnote}}
\pagenumbering{gobble}
\thispagestyle{empty}

\vspace*{2cm}

\begin{center}

{\LARGE\bfseries
Optimization on Manifolds: A Symplectic Approach
}\\[2.0cm]

{\bfseries
Guilherme Fran\c ca,${}^{\!a}$\footnote{\mailto{guifranca@gmail.com}}~%
Alessandro Barp,${}^{\!bc}$\footnote{\mailto{ab2286@cam.uk}} \\[.5em]
Mark Girolami,${}^{\!bc}$\footnote{\mailto{mgirolami@turing.ac.uk}}~%
and Michael I. Jordan${}^{a}$\footnote{\mailto{jordan@cs.berkeley.edu}}
}\\[1cm]

${}^{a}${\itshape University of California, Berkeley, USA}\\[.2em]
${}^{b}${\itshape University of Cambridge, Cambridge, UK}\\[.2em]
${}^{c}${\itshape Alan Turing Institute, London, UK}\\[2cm]

{\bf Abstract}
\end{center}
\vspace{-1em}

Optimization tasks are crucial in statistical machine learning. Recently, there has been great interest in leveraging tools from dynamical systems to derive accelerated and robust optimization methods via suitable discretizations of continuous-time systems. However, these ideas have mostly been limited to Euclidean spaces and unconstrained settings, or to Riemannian gradient flows. In this work, we propose a dissipative extension of Dirac’s theory of constrained Hamiltonian systems as a general framework for solving optimization problems over smooth manifolds, including problems with nonlinear constraints. We develop geometric/symplectic numerical integrators on manifolds that are ``rate-matching,'' i.e., preserve the continuous-time rates of convergence.
In particular, we introduce a
dissipative RATTLE integrator able to achieve optimal convergence
rate locally.
Our class of (accelerated) algorithms are not only simple and efficient but also applicable to a broad range of contexts.

\vfill
\newpage
\tableofcontents
\newpage

\renewcommand*{\thefootnote}{\arabic{footnote}}
\setcounter{footnote}{0}
\pagenumbering{arabic}
\setcounter{page}{1}

\section{Introduction}

We are interested in constructing a general framework for solving optimization problems of the form
\be \label{opt_problem}
\min_{ \qq \in \Q} f(\qq) ,
\ee
where $\Q$ is a smooth manifold, called \emph{configuration manifold}, and 
$f : \Q \to \mathbb{R}$ is a smooth function. 
Such problems have  important applications in
machine learning, statistics, and applied mathematics, including
maxcut problems, phase retrieval, linear and nonlinear eigenvalue problems,
principal component analysis, clustering, and dimensionality reduction, to mention a few
examples.
Configuration manifolds  arise from  rank and orthogonality 
constraints, leading to nonconvex problems. 

The geometry of  $\Q$ is  usually specified by a Riemannian metric. 
However, for  practical reasons, e.g., to avoid computing affine connections
or parallel transports,  it is  convenient to embed $\Q$ into a higher
dimensional manifold, which can be taken to be the Euclidean space $\mathbb{R}^n$. 
The problem can thus be  redefined as minimizing $f$ subject to a set of independent constraints,
\be \label{constraints}
\min_{\qq \in \mathbb{R}^n} f(\qq) \quad \mbox{s.t.} \quad
\gv(\qq) = \bm{0}, 
\ee
where $\gv \equiv (\g_1, \dotsc,\g_m) : \mathbb{R}^{n} \to \mathbb{R}^m$, 
$\psi_a : \mathbb{R}^n \to \mathbb R$, $a=1,\dotsc,m$. 
The configuration manifold is   a $d$-dimensional ($d = n-m$) submanifold  embedded into $\mathbb{R}^n$, namely
$\Q \equiv \{ \qq \in \mathbb{R}^n \, | \, \gv(\qq) = \bm{0} \}$.
Whitney's embedding 
theorem ensures that any smooth real $d$-dimensional manifold
embeds into $\mathbb{R}^{2d}$. Also,  Nash's embedding theorem ensures 
that every  Riemannian manifold can be isometrically embedded 
into $\mathbb{R}^n$ for sufficiently large $n$. Therefore, 
there is no loss of 
generality in this approach. 

To solve problem \eqref{constraints}
we need to  specify
a  dynamics whose trajectories asymptotically converge to critical points, i.e.,
satisfy the Karush–Kuhn–Tucker (KKT) conditions.
To this end, we  propose a \emph{dissipative} extension
of Dirac's theory of constrained Hamiltonian systems \cite{Dirac:1950,Dirac:2001,Henneaux:1994}.
The dynamics is  of second order and  naturally accelerated
in a physical sense. We indeed prove that such a system 
is asymptotically stable and achieves
optimal convergence rate near isolated critical points.
Furthermore, we  generalize problem \eqref{constraints} over a Riemannian manifold
$\M$ under equality and  inequality constraints,
\be \label{constraints2}
\min_{\qq \in \M} f(\qq) \quad \mbox{s.t.} \quad \gv(\qq) = \bm{0}, 
\quad \hhv(\qq) \le \bm{0},
\ee
 and propose a suitable dynamics
for it.

Our approach is different, and in a sense complementary, to that traditionally
found in the optimization literature 
\cite{Absil:2008,Boumal:2023,Boumal:2022,Sra:2016,Pymanot:2016,Ahn:2020,
Alimisis:2021}.
These  approaches invariably  
rely on the Riemannian geometry of $\Q$,
employing (approximations of) the geodesic or gradient flow.
This  is valid once a  Riemannian
metric is  specified. In practice,
 however, this can only be done efficiently 
  for a handful of manifolds with invariant Riemannian metrics. Moreover,  geodesic computations take place on 
the \emph{tangent bundle} $T \Q$, having
the undesired effect of moving states off the manifold;  some
 projection back to $\Q$ is often necessary, which can lead to numerical inefficiencies.
Alternatively,
the \emph{canonical way} to define a dynamics on  $\Q$ is
through a \emph{Hamiltonian system}, whose phase space is instead the 
\emph{cotangent bundle} $T^* \Q$. 
This is because
the cotangent bundle of \emph{any}
smooth manifold is itself a \emph{symplectic manifold}; 
a dynamics that preserves the symplectic structure must be 
 Hamiltonian \cite{Berndt:2000}. In other words, a Hamiltonian system naturally respects and is adapted to the  geometry of the configuration manifold.
Furthermore, through an embedding via constraints, we do not  need to rely
on the Riemannian metric of $\Q$.

Relations between Hamiltonian systems and optimization on Euclidean spaces 
have been attracting great interest; see, e.g., \cite{Wibisono:2016,Franca:2020,Betancourt:2018,muehlebach:2021,Bravetti:2019,Franca:2021b,Franca:2023} and references therein.
However,  connections with Hamiltonian systems over general smooth manifolds, or
with constrained Hamiltonian systems, have not yet been explored.
It is the goal of this paper to fill this important gap.

Once a suitable  dynamics has been identified, one needs 
to introduce discretizations that are  stable and
retain  its important properties. 
Symplectic integrators are the preferred choice for simulating Hamiltonian 
systems due to their long term stability, for exactly preserving
the symplectic structure, and for having close energy conservation
\cite{Benettin:1994,McLachlan:2002,Leimkuhler:2004,McLachlan:2006,Hairer:2010}.
However, symplectic integrators were developed for conservative systems.
Recently, an extension of symplectic integrators to dissipative settings, yielding methods
that closely preserve continuous-time rates,
was proposed 
\cite{Franca:2021}. This framework again 
applies to Euclidean and unconstrained settings.
Here we further extend these ideas to arbitrary smooth manifolds and for constrained
cases, thereby significantly enlarging the range of applications that
include problems not only in machine learning and optimization but also in molecular dynamics, control theory, complex systems, and statistical physics.
We therefore provide a first principles derivation---based on symplectic geometry and backward error analysis---of manifold/constrained optimization methods that emulate dissipative Hamiltonian systems. 
Such discretizations preserve the continuous-time 
rates of convergence via  the preservation of a \emph{presymplectic} structure.

The main contributions of this paper are shortly summarized as follows:
\begin{itemize}
\item We provide a general framework based on continuous-time systems
for optimization on smooth
manifolds as well as for problems with nonlinear equality/inequality constraints.
This is done by a dissipative extension of Dirac's theory of constrained
Hamiltonian systems.
\item We extend the theory of symplectic integrators to construct
practical optimization algorithms to this manifold/constrained setting that
are ``rate-matching.''
\item We introduce a system that satisfies the KKT conditions (on curved
spaces) and have optimal rates of convergence locally,
i.e., accelerated rates sufficiently close to critical points. 
\item  We derive a simple and efficient algorithm that consists of a dissipative
generalization of the famous  RATTLE integrator from molecular
dynamics (our framework  allows a variety of other methods to be 
obtained as well, and we provide extensions in the Appendix).
\end{itemize}

The outline of the paper is the following.
In Sec.~\ref{sec:general},
we introduce dissipative and constrained Hamiltonian systems over
Riemannian manifolds, emphasizing their \emph{symplectification}, which
plays a fundamental role for introducing our class of presymplectic integrators.
Such discretizations are shown to generalize important properties of symplectic integrators (shadowing).
In Sec.~\ref{sec:diss_geo}, we propose a dissipative geodesic equation
with equality constraints, and show stability and local (accelerated) convergence rate
results.
In Sec.~\ref{sec:gen_kkt}, we argue that this can be extended to inequality
constraints, leading to a general dynamics in consistency with  KKT conditions on Riemannian manifolds.
In Sec.~\ref{sec:diss_geo_rattle}, we state an extension of RATTLE for constrained optimization, with
numerical experiments shown in Sec.~\ref{sec:numerical}.
We conclude in Sec.~\ref{sec:conclusion}. 
Note that this paper assumes sufficient background on differential geometry and
Hamiltonian systems. Nevertheless, we provide a quick review of basic concepts in the Appendix,
which also contains all the technical proofs, omissions, and several additional
results not stated in the main part of the paper.

\textit{Notation.}
We refer to vectors in boldface, $\qq \in \mathbb{R}^n$, and its components
are denoted by $\q^i$, for $i=1,\dotsc,n$ (upper indices).
The dual or covector $\pp \in \mathbb{R}^n$ has components 
$\p_i$, for $i=1,\dotsc,n$ (lower indices). The metric tensor $g$ of type $(2,0)$ has
components $g_{ij}$, and its inverse $g^{-1}$, 
which is a tensor of type $(0,2)$,
has components  $g^{ij}$.
We interchangeably refer to vector valued functions as
$f(\qq)$ or $f(\q^i)$. 
The gradient $\nabla f(\qq)$ has components $\partial_i f$ (covector), where
$\partial_i \equiv \partial / \partial \q^i$.
We use Einstein's 
summation  on repeated upper and lower indices, e.g.,
$q^i p_i \equiv \sum_{i=1}^n q^i p_i$, $\lambda^a\g_a \equiv \sum_{a=1}^m \lambda^a \g_a$, $g^{ij}p_j \equiv \sum_{j=1}^n g^{ij} p_j$, etc.
The symbol $\wedge$ denotes the exterior product between differential forms.
For simplicity,  we often omit  variable dependencies, e.g., the function
$H(t, q^i, p_i)$ may be denoted simply by $H$, and $\g_a(\q^i)$ by $\g_a$.
Other necessary terminologies will be introduced along the paper.

\section{General Framework}
\label{sec:general}

\subsection{Nonconservative and constrained Hamiltonian systems}

Let $\M$ be an $n$-dimensional Riemannian manifold
with local coordinates $\q^i $, and let
 $ \p_i $ be their  conjugate momenta ($i=1,\dotsc,n$).  We thus
 have local coordinates  
$ ( \q^i, \p_i)  \in \mathbb{R}^{2n}$  on the cotangent bunddle $T^*\M$.
Let $H(t, \qq, \pp)$ be an unconstrained Hamiltonian function describing
some physical system, allowed to be
explicit time dependent so as to account for \emph{dissipation}.
Consider a set of so-called \emph{primary constraints}, 
$\g_a(t, \qq, \pp)$,  $a=1,\dotsc,m$ \cite{Dirac:1950,Dirac:2001,Henneaux:1994}---%
in this section we allow the constraints to be more general than actually needed for 
problem \eqref{constraints}.
A mechanical system subject to these
constraints can be defined by the  total Hamiltonian
\be \label{H1}
H_{\text{total}}  \equiv
H(t, \qq, \pp) + \lambda^a \g_a(t, \qq, \pp) ,
\ee
where $\lambda^a$ are  Lagrange multipliers.
The constraints define the configuration
manifold $\Q$, which is embedded into $\M$. 
Hamilton's equations,
$\dot{q}^i = \partial H_{\text{total}} / \partial p_i$ and
$\dot{p}^i = - \partial H_{\text{total}} / \partial q^i$, 
yield 
\be \label{ham1}
\dot{\q}^i \approx 
\dfrac{\partial H}{\partial \p_i} +  \lambda^a \dfrac{\partial \g_a}{\partial p_i} , \qquad
\dot{\p}_i \approx -\dfrac{\partial H}{\partial \q^i} - 
 \lambda^a \dfrac{\partial \psi_a}{\partial \q^i} .
\ee
We use  Dirac's notation \cite{Dirac:1950,Dirac:2001,Henneaux:1994}, where  $\approx$ means ``weak equality'' that holds only on the constraint surface $\g_a(t, \qq, \pp) = 0$, 
i.e., it is implicit that the constraint condition is obeyed.
The above system is \emph{nonconservative}, $d H_{\text{total}}/ dt =
\partial H_{\text{total}} / \partial t$,   yielding
\be \label{noncons}
\dfrac{d H}{dt} \approx \dfrac{\partial H}{\partial t}
\ee
if we assume that $\partial \g_a / \partial t \approx 0$.
This is an important ingredient, and  holds  if $\g_a$ has a factorized form, 
$\g_a(t,\qq,\pp) \mapsto h(t) \g_a(\qq, \pp)$, for some function $h(t)$, 
as will be used shortly---%
note that we always have $d \g_a / dt \approx 0$ since the constraints are  
obeyed at all times.
 If we explicitly solve for the
Lagrange multipliers $\lambda^a$, which we assume are unique as usual, the
differential algebraic equations \eqref{ham1} 
reduce to ordinary differential equations for
$2n$ degrees of freedom, which  are not all independent, i.e., 
the phase space of the system is a submanifold
$T^*\Q \subset T^*\M$ of codimension $2(n-m)$.

\subsection{Symplectification}

\begin{figure}
\centering
\includegraphics[scale=1.05]{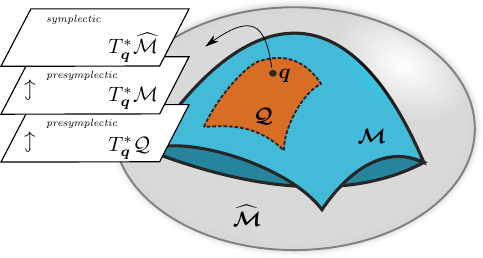}
\caption{\label{fig:embedding} Illustration of the symplectification.}
\end{figure}

We now consider a \emph{symplectification}  \cite{Berndt:2000},
whereby the phase space of the above system can be embedded
into a higher dimensional \emph{symplectic manifold}---%
this  will play a fundamental role for introducing geometric integrators.
Thus, we promote time to a new coordinate, $q^0  \equiv t$, and 
let the Lagrange multipliers $\lambda^a$ become  dynamical variables.  
We also introduce their conjugate momenta $p_0$ and $\pi_a$, respectively.
Define $Q^I \equiv (q^\mu, \lambda^a)$ and $P_I \equiv (p_\mu, \pi_a)$, for
$I = 0,\dotsc,n+m$.
We can view this as a sequence of embeddings,  
\be
\Q^{n-m} \hookrightarrow \M^n \hookrightarrow \widehat{\M}{}^{n+m+1},
\ee
where $\widehat{\M}{}^{n+m+1} \equiv \M^{n} \times \mathbb{R}^m \times \mathbb{R}$.
The associated cotangent bundle $T^*\widehat{\M}$ has dimension
$2(n+m+1)$ and   local coordinates $( Q^I, P_I)$.
On this \emph{extended phase space}, we can define a generic Hamiltonian
$\mathscr{H}(\bm{Q}, \bm{P})$  with standard Hamilton's equations,
\be\label{ham22}
\dfrac{d Q^I}{ds} = \dfrac{\partial \mathscr{H}}{ \partial P_I}, \qquad
\dfrac{d P_I}{ds} = - \dfrac{\partial \mathscr{H}}{ \partial Q^I}.
\ee
This  system is 
conservative, $d \mathscr{H} / ds = 0$, since $\mathscr{H}$ does not depend
explicitly on time---we denote the ``new''
time parametrization by $s$.
Now, suppose we choose this Hamiltonian to have the  form
\be  \label{spe_ham}
\mathscr H\big(\bm{Q}, \bm{P} \big) \equiv H_{\text{total}}\big(q^0, \qq, \lambdav,\pp \big) + p_0 .
\ee
The equations of motion \eqref{ham22} then become (see 
Fig.~\ref{fig:embedding} for an illustration
of the symplectification procedure)
\begin{subequations} \label{ham2}
\begin{align}
\dfrac{d q^0}{ds} &= 1, \label{g1} \\
\dfrac{d p_0}{ds} &= - \dfrac{\partial H }{ \partial q^0} -  
\lambda^a \dfrac{\partial \g_a}{\partial q^0},  \label{g2} \\
\dfrac{d q^i}{ds} &= \dfrac{ \partial H}{\partial p_i} + 
\lambda^a \dfrac{\partial \g_a}{\partial p_i}, \label{g3} \\
\dfrac{d p_i}{ds} &= - \dfrac{ \partial H }{\partial q_i} -
\lambda^a  \dfrac{\partial \g_a}{\partial q^i},  \label{g4} \\
\dfrac{d \lambda^a}{ds} &= 0, \label{g5} \\
\dfrac{d \pi_a}{ds} &= - \psi_a. \label{g6}
\end{align}
\end{subequations}
We can  enforce the constraints by setting $\pi_a = 0$, so that
Eq.~\eqref{g6} becomes $\psi_a(q^0, \qq, \pp) = 0$.
Eq.~\eqref{g5} simply says that the Lagrange multipliers do not change
on the constraint surface.
Eq.~\eqref{g1} yields $q^0 = t = s$ (by setting the integration constant to zero).
Eq.~\eqref{g2} becomes
$dp_0 / dt = -\partial H_{\text{total}} / \partial t = d H_{\text{total}} / dt$, i.e.,
$p_0(t) = - H_{\text{total}}(t) \approx -H(t)$, up to a constant that can be chosen to be zero.
Thus, $p_0$ is  simply the value of the original Hamiltonian on the constraint surface.
Finally, Eqs.~\eqref{g3} and \eqref{g4} yield the original equations
of motion \eqref{ham1}.
Therefore, we have embedded the degrees of freedom of system \eqref{ham1} into
the phase space of a higher dimensional \emph{conservative} Hamiltonian system
\eqref{ham22}.
The original nonconservative and constrained system is recovered from 
\eqref{spe_ham} under the ``gauge fixing''
\be
\label{gauge}
\pi_a = 0, \qquad s = q^0 = t, \qquad p_0(t) = -H_{\text{total}}(t)  \approx - H(t) .
\ee
The phase space of the original system 
lives on a  hypersurface $\partial \mathcal{S}$ embedded into $T^* \widehat{\M}$,  defined by 
\be \label{hyper}
\mathscr{H}\big\vert_{\partial \mathcal{S}}  
\approx H(q^0, \qq, \pp) + p_0 \approx  0.
\ee

\begin{definition}[See \cite{Berndt:2000}]
A smooth manifold of dimension $2n$ endowed with a closed and nondegenerate 
2-form $\Omega$ is called a \emph{symplectic manifold}, and $\Omega$ is called
a \emph{symplectic form}.
A smooth manifold of dimension $2n + \bar{n}$, $\bar n >0$, endowed
with a closed 2-form $\omega$ of rank $2n$ that  is 
degenerate is
called a \emph{presymplectic manifold}, and $\omega$ is called a \emph{presymplectic form}.
\end{definition}

The  Hamiltonian
system~\eqref{ham22} has a closed and nondegenerate 
symplectic 2-form given by
\be \label{Omb}
\begin{split}
 \Omega \equiv dQ^I \wedge dP_I 
= dq^0 \wedge p_0 + dq^i \wedge dp_i + d\lambda^a \wedge d\pi_a  ,
\end{split}
\ee
which  is preserved, i.e., the Lie derivative along the flow vanishes,
 $\mathcal{L}_{\mathscr{H}} \Omega = 0$.
 Thus, $(T^*\widehat{\M}, \Omega)$ is a \emph{symplectic  manifold}.
 However, under the gauge fixing \eqref{gauge}, we recover 
 the nonconservative and constrained system \eqref{ham1} with a \emph{restricted
presymplectic} 2-form
\be \label{om}
\omega \equiv \Omega\vert_{\partial \mathcal{S}} = dq^i \wedge dp_i ,
\ee
which has rank $2n$ over a space of dimension 
$2n + 2m + 2$. It therefore follows that $\omega$ is degenerate, and 
the phase space $(T^*\Q, \omega)$ of system 
\eqref{ham1}  is  a \emph{presymplectic manifold}.
For conservative and constrained Hamiltonian systems such a presymplectic
structure has already being noted \cite{Gotay:1978}, as well as for 
dissipative but unconstrained cases 
\cite{Franca:2021}. The above discussion incorporates  both situations.
We say that  system \eqref{ham1} admits a \emph{symplectification}
since its phase space can be recovered by a restriction of a higher dimensional
symplectic phase space.

\subsection{Presymplectic integrators}

As we have seen, a conservative Hamiltonian system
\eqref{ham22} has a symplectic form $\Omega$ that is constant.
Let $\Phi_h :  T^* \widehat{\M} \to T^*\widehat{\M}$ be a \emph{symplectic integrator}
for  system \eqref{ham22}, where $h > 0$ is the discretization
step size such that $s_\ell = h \ell$, for iterations $\ell = 0,1,\dotsc$ ($s$ denotes
time).
A symplectic integrator is by definition a discretization that exactly
preserves the symplectic form
\cite{Benettin:1994,McLachlan:2002,Leimkuhler:2004,McLachlan:2006,Hairer:2010},
i.e., $\Omega(s_\ell) = \Phi_h^\ell \circ \Omega(0) = \Omega(0)$, where
$\Phi_h^\ell$ denotes $\ell$ iterations of the map $\Phi_h$.
The numerical integrator is said to be of order $r > 1$ if the global error
in approximating the continuous-time trajectory is $\bigO(h^r)$ \cite{Hairer:2010}.

\begin{theorem}[See \cite{Franca:2021}] \label{thm_local_ham}
A dynamical system with phase space $T^*\widehat{\M}$ perserves the symplectic form
$\Omega$ if and only if it is (locally) a conservative Hamiltonian 
system \eqref{ham22}.
\end{theorem}
As a consequence, a symplectic integrator admits 
a (local) Hamiltonian which is a perturbation in terms of the step size,
$\widetilde{\mathscr{H}} = \mathscr{H} + h^r \Delta \mathscr{H}_1 + \dotsm$,
called  \emph{shadow Hamiltonian}
\cite{Benettin:1994}. This allows one to study the numerical method purely from
a continuum  analysis. The existence of a shadow Hamiltonian is
key in explaining the success of symplectic integrators. 
(Other types of integrators that are not symplectic 
do not admit a shadow Hamiltonian.)
We now introduce similar notion 
to symplectic integrators but for nonconservative and constrained systems.

\begin{definition}\label{presymp_def}
A numerical map $\phi_h : T^*\Q \to T^*\Q$ is said to be a \emph{presymplectic
integrator} to the nonconservative and constrained Hamiltonian system
\eqref{ham1} if it is obtained by a reduction, i.e., by the gauge fixing
\eqref{gauge}, of a symplectic integrator $\Phi_h$ for its symplectification, namely
system \eqref{ham22}/\eqref{spe_ham}.
\end{definition}

Thus, presymplectic integrators are essentially constructed from 
symplectic integrators. The crucial difference is that they preserve the presymplectic
form $\omega$, which is no longer constant.
Given a conservative Hamiltonian, obtained from a non-singular Lagrangian, 
one can construct the exact discrete
Lagrangian which solves the Hamilton-Jacobi
equation \cite{Marsden:2001}. Integrators obtained from a discrete
Lagrangian are symplectic and admit a \emph{globally} 
defined shadow Hamiltonian.
Such integrators are called \emph{variational integrators}. 

\begin{assumption}\label{assump_var}
We assume the base symplectic integrator $\Phi_h$ is a variational integrator \cite{Marsden:2001}.
\end{assumption}

With these ingredients, we
state our next result that
extends one of the main results of \cite{Franca:2021} to arbitrary smooth
manifolds $\M$ and for constrained cases.
\begin{theorem}\label{thm_ham_preserv}
Let $\phi_h$ be a presymplectic integrator of order $r$ for the
nonconservative and constrained Hamiltonian system \eqref{ham1}, whose
true flow is denoted by $\varphi_t$. Then this method preserves the
time-varying Hamiltonian up to a bounded error,
\be\label{ham_preserv}
H \circ \phi_{h}^\ell = H \circ \varphi_{t_\ell} + \bigO(h^r),
\ee
where $t_\ell = h \ell = \bigO(h^r e^{c/h})$ is the simulation time, with 
some constant $c>0$, and $\ell = 0,1,\dotsc$.
Moreover, the method admits a shadow Hamiltonian
$\widetilde{H} = H + h^r \Delta H_r + \dotsm$ that is globally defined.
\end{theorem}
\vspace{-.5em}
\begin{proof}[Proof sketch]
The basic idea involves using the symplectification \eqref{spe_ham},
allowing us to rely on standard results for symplectic integrators
ensuring $\mathscr{H}\circ \Phi_h^\ell = \mathscr{H} \circ \widehat{\varphi}_{s_\ell} + \bigO(h^r)$, where $\Phi_h$ is the base variational integrator and
$\widehat{\varphi}_{s_\ell}$ is the true flow of \eqref{spe_ham}.
We then apply the gauge fixing \eqref{gauge} on the numerical map to
obtain \eqref{ham_preserv}. The shadow Hamiltonian for $\phi_h$  follows
under Assumption~\ref{assump_var}. 
\end{proof}

\section{Dissipative Geodesic under Constraints}
\label{sec:diss_geo}

Consider the total Hamiltonian
\be\label{hamM}
H_{\text{total}} = \dfrac{1}{2} e^{-\eta(t)} g^{ij}(\qq) \p_i \p_j + 
e^{\eta(t)} f(\qq) + e^{\eta(t)} \lambda^a \g_a(\qq) ,
\ee
where $\eta(t) > 0$ is a generic dissipation function, and $g_{ij}$ is the Riemannian
metric of $\M$. Note that we choose the constraints in \eqref{H1} to have
the factorized form $\g_a(t, \qq, \pp) \mapsto e^{\eta(t)} \g_a(\qq)$, and now $\g_a$ depends only on the position (holonomic).
Hamilton's equations for \eqref{hamM} can be reduced to 
\be\label{dissip_geo}
\ddot{q}^i + \Gamma\indices{^i_j_k} \dot{q}^j \dot{q}^k + \dot{\eta}(t) \dot{q}^i
\approx - g^{ij} \partial_j f - g^{ij} \lambda^a \partial_j \g_a ,
\ee
where $\Gamma\indices{^i_j_k}$ are the Christoffel symbols.
This is  a \emph{dissipative geodesic equation} subject to
constraints $\gv(\qq) = \bm{0}$---note that if $\dot{\eta} = 0$, $f=0$ 
and $\g_a = 0$ we recover a standard geodesic over $\M$. 
Suppose that for such a general 
system one is able to prove a rate of convergence. Then a presymplectic
integrator closely reproduces this rate as follows.

\begin{corollary}\label{thm_rate_preserv}
Consider system \eqref{dissip_geo} over a Riemannian
manifold $\M$ and  subject to constraints.
Let $\phi_h$ be a presymplectic integrator of order $r$ (Definition~\ref{presymp_def}). 
Assume this numerical integrator has a Lipschitz constant $L_\phi > 0$. 
Then such an integrator preserves the convergence rates in the sense that
\be\label{preserv_rate}
\underbrace{f \circ \phi_h^\ell - \min f}_{\textnormal{discrete-time rate}} = \underbrace{f\circ\varphi_{t_\ell} -\min f}_{\textnormal{continuous-time rate}}  
+ \underbrace{\bigO(h^r e^{-\eta(t_\ell)})}_{\textnormal{small error}} \,,
\ee
provided $e^{L_\phi - \eta(t_\ell)} < \infty$ and $t_\ell = \bigO(h^r e^{c/h})$
for some constant $c > 0$.
\end{corollary}
\begin{proof}[Proof sketch]
The result is an immediate consequence of Eq.~\eqref{ham_preserv} together
with the Lipschitz condition for the numerical map (which is standard).
\end{proof}

This result shows that presymplectic integrators are able to closely preserve
the continuous-time rates of convergence in general.
The error term can be exponentially small,  thus  negligible in practice,
and the simulation time exponentially large, 
not incurring practical limitations.
Next, we show some stability analysis results for system \eqref{dissip_geo}.

\begin{theorem}\label{thm_kkt1}
The pair $(\qq_\star, \lambdav_\star) $ is a critical point of system
\eqref{dissip_geo}  only if it satisfies the KKT conditions
for the optimization problem
\be\label{opt_problem2}
\min_{\qq \in \M} f(\qq) \quad \textnormal{s.t.} \quad
\gv(\qq) = \bm{0},
\ee
namely
\be\label{kkt1}
\partial_i f(\qq_\star) +  \lambda_\star^a \partial_i \g_a(\qq_\star) = 0, 
\qquad \g_a(\qq_\star) = 0,
\ee
for $i=1,\dotsc,n$ and  $a=1,\dotsc,m$.
\end{theorem}
\begin{proof}[Proof sketch]
This  follows immediately by writing Eq.~\eqref{dissip_geo} in
first order form.
\end{proof}

\begin{theorem}\label{thm_stability}
The system \eqref{dissip_geo} is stable around an isolated minimizer
of problem \eqref{opt_problem2}.
In addition, if the damping is constant, $\eta(t) = \gamma t$ with $\gamma = \text{const.} > 0$,
then the system is asymptotically stable. 
\end{theorem}
\begin{proof}[Proof sketch]
This is obtained by considering the Lyapunov function
$\mathcal{E} = (1/2) g^{ij}(\qq) p_i p_j + f(\qq) + \lambda^a \g_a(\qq)$,
which is the mechanical energy, and showing that
$\dot{\mathcal{E}} \le 0$. When the damping is constant, asymptotic
stability follows from LaSalle's invariance principle.
\end{proof}

These last two results show that system \eqref{dissip_geo} is able to solve problem
\eqref{opt_problem2} under suitable conditions.
We are also interested in knowing how fast it converges to the solution.
We provide a local result.

\begin{theorem}\label{thm_rate}
In a sufficiently small neighborhood of an isolated minimizer $\qq_\star$ of problem
\eqref{opt_problem2}, the system \eqref{dissip_geo} with a constant damping $\eta = \gamma t$
such that $\gamma \le 2 \sqrt{\omega_{\text{min}}}$, and initial conditions
$\qq_0 \equiv \qq(0)$ and $\dot{\qq}(0) = \bm{0}$ obeying the constraints, 
has a convergence rate of
\be \label{rate}
\| \qq(t) - \qq_\star \|^2 \le e^{ -2 \sqrt{ \omega_{\text{min}} } \, t } 
\| \qq_0 - \qq_\star\|^2 ,
\ee
where $\omega_{\text{min}} > 0$ is the smallest eigenvalue of 
the projected Hessian
$\big(g^{-1} \mathcal{P} \nabla^2 f \mathcal{P}^T g^{-1} \big)\big\vert_{\qq_\star}$
at the critical point, where
$\mathcal{P}$ is the projection operator to the constraint surface defined
in Eq.~\eqref{proj2}.
\end{theorem}

This result implies that that one can obtain discretizations able to achieve
 \emph{optimal rates} locally. 
To see this, suppose we have a discretization with step size 
$h > 0$, $t_\ell = h \ell$, for iterations $\ell = 0,1,\dotsc$.
Restoring a mass $m$ into \eqref{dissip_geo} (see Eq.~\eqref{ode_mass}
in the Appendix) the rate \eqref{rate} changes to $e^{
-2\sqrt{\omega_{\text{min}}/m}t}$.
Choosing $m=h$, step size
\be \label{hC2}
h = C^2 / \omega_{\text{max}} ,
\ee
where $\omega_{\text{max}}$ is the largest eigenvalue of 
the projected Hessian from Theorem~\ref{thm_rate}, and
$C$ is related to the numerical stability of the  integrator,
the rate \eqref{rate} becomes
\be \label{rateQ}
\| \qq_{\ell} - \qq_\star \|^2 \le e^{-2 C \sqrt{Q^{-1}} \ell} \| \qq_0 - \qq_\star\| ,
\ee
provided the error in the discretization can be neglected compared to this
rate---which  is  the case for presymplectic integrators according to Eq.~\eqref{preserv_rate}. 
Above, $Q \equiv \omega_\text{max} / \omega_{\text{min}}$ is the \emph{condition
number} of the problem.  Interestingly, for the presymplectic integrator we 
introduce in the next section,  $C = 2$, in which case the rate \eqref{rateQ}
achieves precisely
the well-known lower bound (optimal rate) for smooth, strongly convex,
but \emph{unconstrained} problems on $\mathbb{R}^n$,
namely \cite{Nesterov:2018}
\be
\label{lower_bound}
\| \qq_\ell - \qq_\star \|^2 \ge \left( \dfrac{\sqrt{Q} - 1}{\sqrt{Q} + 1} \right)^{2\ell} \| \qq_0 - \qq_\star\|^2 \simeq e^{-4 \sqrt{Q^{-1}} \ell} \| \qq_0 - \qq_\star \|^2.
\ee
Thus, suitable discretizations of system \eqref{dissip_geo} are able
to achieve this optimal/accelerated lower bound at least locally 
(or get very close to it
depending on $C$), and on more general settings than previously considered.

\section{General KKT Conditions on Manifolds}
\label{sec:gen_kkt}

Consider adding inequality constraints to problem \eqref{opt_problem2}, 
\be\label{opt_problem3}
\min_{\qq \in \M} f(\qq) \quad \mbox{s.t.} \quad
\gv(\qq) = \bm{0}, \quad \hhv(\qq) \le \bm{0} ,
\ee
where $\hhv = (\hh_1,\dotsc, \hh_{\overline{m}})$. 
Inside the feasible region,  inequality constraints are 
\emph{inactive}. They only play a role and become \emph{active} on the boundary of the region. 
Effectively, inequality constraints  thus behave  as equality constraints but 
they can be switched ``on/off.'' 
Let us first focus on a single inequality constraint $\hh_b$ ($b=1,\dotsc,\overline{m}$) with
associated Lagrange multiplier $\mu^b$. We have either 
$\hh_b(\qq) < 0$, in which case we can set $\mu^b = 0$ since the constraint
is inactive, or $\hh_b(\qq) = 0$, in which case $\mu^b \ne 0$. We thus have (with no summation over $b$ implied)
\be\label{slackness}
\mu^b \cdot \hh_b(\qq) = 0 .
\ee
Geometrically, the constraint is active if it ``pulls'' in the opposite of
$f$'s descent direction, i.e., $(-\partial_i f)(-\partial^i \hh_b) \le 0$.
Thus, at a given point $\qq$, if $\partial_i f \partial^i \hh_b \le 0$ then 
the constraint $\hh_b$ is active, and otherwise it is inactive.
A critical point must obey
$ \partial_j f(\qq_\star) + \mu^c_\star \partial_j \phi_c(\qq^\star)  = 0$, for
$c = 1,\dotsc,\overline{m}$. Contracting this last expression with $\partial_i \hh_b(\qq_\star)$
we conclude that
\be\label{Dab}
\mu_\star^c \partial_i \hh_b(\qq_\star)\partial^i \hh_c(\qq_\star) \ge 0,
\ee
where we used  $\mu_\star^c = 0$ for inactive constraints.
Noticing that $\partial_i \hh_b \partial^i \hh_c = \big( \partial_{\qq}\hhv \, \partial_{\qq} \hhv^T \big)_{bc}$ are entries
of a positive semidefinite matrix, this relation implies that
\be\label{dual_feasibility}
\mu_\star^b \ge 0 \qquad (b=1,\dotsc,\overline{m}).
\ee

Now, consider the Hamiltonian system
\be\label{hamM2}
H_{\text{total}} = \dfrac{1}{2} e^{-\eta(t)} g^{ij}(\qq) \p_i \p_j + 
e^{\eta(t)} f(\qq) + e^{\eta(t)} \lambda^a \g_a(\qq) +
e^{\eta(t)} \mu^b \hh_b(\qq),
\ee
which yield the equations of motion
\begin{subequations}\label{dissip_geo2}
\begin{align}
\ddot{q}^i + \Gamma\indices{^i_j_k}(\qq) \dot{q}^j \dot{q}^k + \dot{\eta}(t) \dot{q}^i
&= - g^{ij}(\qq) \left[
\partial_j f(\qq) + \lambda^a \partial_j \g_a(\qq) +\mu^b \partial_j \hh_b(\qq) \right] , \label{dineqgeo1}\\
0 &= \g_a(\qq) \qquad (a=1,\dotsc,m), \label{dineqgeo2} \\
0 &\ge \hh_b(\qq) \qquad (b=1,\dotsc,\overline{m}) . \label{dineqgeo3}
\end{align}
\end{subequations}
Importantly,  it is implicit that  $\mu^b(t) =0$ whenever
$\hh_b(\qq(t)) < 0$ (inactive), so that the system becomes unconstrained with
respect to $\hh_b$ in such case. Therefore, Eq.~\eqref{slackness} is always satisfied
during the evolution of the system. The condition \eqref{dineqgeo3} is effectively an
equality, in the same way as \eqref{dineqgeo2}.
The argument used in obtaining condition \eqref{Dab} at stationarity is still valid, which
implies \eqref{dual_feasibility}. 
We thus have:
\begin{itemize}
\item Critical points of system \eqref{dissip_geo2} obey the KKT conditions
(see Theorem~\ref{thm_gen_kkt} in the Appendix)
\begin{subequations}\label{gen_kkt}
\begin{align}
\partial_j f(\qq_\star) + \lambda_\star^a \partial_j \g_a(\qq_\star) + \mu_\star^b \partial_j \hh_b(\qq_\star) &= 0, 
\label{gen_kkt1} \\
\g_a(\qq_\star) = 0, \quad \hh_b(\qq_\star) &\le 0, 
\label{gen_kkt2} \\
\mu_\star^b \cdot \hh_b(\qq_\star) &= 0, 
\label{gen_kkt3} \\
\mu_\star^b &\ge 0.
\label{gen_kkt4}
\end{align}
\end{subequations}
\item Theorems~\ref{thm_stability} and \ref{thm_rate} also remain true
in this case. The reason is because an inequality constraint $\hh_b$ only participates
in the dynamics when it is active, but then it behaves as an
equality constraint, $\hh_b(\qq) = 0$ when $\mu^b \ne 0$ 
(see the Appendix for  details). 
\end{itemize}

\section{Dissipative  RATTLE for Constrained Optimization}
\label{sec:diss_geo_rattle}

The system \eqref{dissip_geo2} is very general as it incorporates constraints over 
an arbitrary Riemannian manifold $\M$. A numerical simulation requires computing $\Gamma\indices{^i_j_k}$, which involves  derivatives
of the metric  $g_{ij}$. Recalling the discussion
in the introduction, it is convenient to use an
embedding into $\M \equiv \mathbb{R}^n$, where  $g_{ij}$ can be taken 
globally as \emph{constant}, so that $\Gamma\indices{^i_j_k} = 0$.
This  simplifies the construction of geometric integrators,
resulting in simple and explicit updates. Thus, here we focus on this
case and propose a presymplectic
integrator for solving the problem
\be\label{opt_problem4}
\min_{\qq \in \mathbb{R}^n} f(\qq) \quad \mbox{s.t.} \quad
\gv(\qq) = \bm{0}, \quad \hhv(\qq) \le 0 .
\ee
This is done by simulating \eqref{dissip_geo2} with $\Gamma\indices{^i_j_k} = 0$
and $\eta(t) = \gamma t$,  $\gamma = \text{const.} > 0$.
Following Definition~\ref{presymp_def}, we derive
a presymplectic integrator of order $r=2$  based on a modern
reformulation of
the famous RATTLE integrator \cite{Andersen:1983,Leimkuhler:1994,Reich:1996,Leimkuhler:2016},
widely used in molecular dynamics.
The derivation is long but straightforward, and involves a few
technical tricks that are fully described in the Appendix.

To incorporate both equality and inequality constraints, define the
augmented vector of constraints
$\bm{\Psi} \equiv (\g_1, \dotsc, \g_m; \hh_1, \dotsc, \g_{\overline{m}})$,
and Lagrange multipliers 
$\bm{\Lambda} \equiv (\lambda_1, \dotsc, \lambda_m; \mu_1, \dotsc, \mu_{\overline{m}})$.
Let  $\mathcal{J}(\qq)$ denote the $(m+\overline{m})\times n$ Jacobian matrix of constraints at point $\qq$,  
defined as
\be \label{Jactot}
\mathcal{J}_{ai}(\qq)  \equiv 
\begin{cases}
\partial \Psi_a / \partial \q^i\big\vert_{\qq} & \mbox{if $\Psi_a(\qq)$ is active}, \\
0 & \mbox{if $\Psi_a(\qq)$ is inactive,}
\end{cases}
\ee
where $i=1,\dotsc,n$ and $a=1,\dotsc,m + \overline{m}$. 
Equality constraints, $\Psi_a = \g_a$, are always active. 
Inequality constraints, $\Psi_a = \hh_a$, must be checked point-wise.
Define also the projection operators
\be\label{proj2}
\mathcal{R}(\qq) \equiv \mathcal{J}(\qq) g^{-1} \mathcal{J}(\qq)^T, \qquad
\mathcal{P}(\qq) \equiv I -  \mathcal{J}(\qq)^T \mathcal{R}(\qq)^{-1} \mathcal{J}(\qq) g^{-1} .
\ee
With these ingredients, we state Algorithm~\ref{dissgeorattle}.

\begin{algorithm}[t]
\caption{\label{dissgeorattle}
\textsc{DissRATTLE} is a presymplectic integrator for 
 problem \eqref{opt_problem4}. The algorithm has one parameter
$\alpha \in (0,1)$ (momentum factor) and step size $h > 0$.
The parameter $\beta = \cosh(-\log \alpha)$ is fixed, and $g\succ 0$ is an
arbitrary  symmetric matrix (preconditioner).}
\begin{algorithmic}[1]
\For{$\ell=0,1,\dotsc$}
\State $\pp_{\ell+1/2} \leftarrow \alpha  \mathcal{P}(\qq_{\ell})\big[ \pp_{\ell}  -
(h/2) \nabla f(\qq_{\ell})\big]$ \label{geo1}
\State $\widetilde{\pp}_{\ell+1/2} \leftarrow  \pp_{\ell+1/2}-(h \alpha /2) \mathcal{J}(\qq_\ell)^T \bm{\Lambda}_\ell$  \label{geo2} 
\State $\qq_{\ell+1} \leftarrow \qq_{\ell} + \beta g^{-1} \widetilde{\pp}_{\ell+1/2}$  \label{geo3}
\For{$a=1,\dotsc,m+\overline{m}$}
\If{$\Psi_a$ is active}
\State $\Lambda_{a,\ell} \leftarrow \Psi_a(\qq_{\ell+1}) = 0$
\Else
\State $\Lambda_{a,\ell} \leftarrow 0$
\EndIf
\EndFor
\State $\pp_{\ell+1} \leftarrow \mathcal{P}(\qq_{\ell+1})\big[ \alpha \widetilde{\pp}_{\ell+1/2} -
(h/2) \nabla f(\qq_{\ell+1}) \big]$ \label{geo5}
\EndFor
\end{algorithmic}
\end{algorithm}


The parameter $\alpha \in (0,1)$  is related to the damping $\gamma$.
Increasing $\alpha$ corresponds to decreasing $\gamma$,
i.e., $\alpha \to 1$ corresponds to $\gamma \to 0$,  and
$\alpha \to 0$ to $\gamma \to \infty$.
The parameter  $\beta \equiv \cosh(-\log \alpha)$ is fixed.
The updates 2--11
must be solved
simultaneously, with the components of  
$\bm{\Lambda}_\ell$ determined by solutions
to the algebraic equations in step 7 when the constraint $\Psi_a$ is active,
and otherwise the Lagrange multiplier is set to zero. 
We wrote component-wise  but in practice these updates can be computed 
vectorially with the aid of any root finding routine  
(when a closed form  solution is unavailable).
This algorithm requires only one gradient computation of $f$ per iteration (the last one
can be reused in the subsequent iteration). 
Finally, instead of a constant $\alpha$ one
is  free to use an adaptive $\alpha_\ell$, which  would be associated to a 
time-dependent damping  $\gamma \mapsto \gamma(t)$.
Conveniently, this method  uses the Euclidean gradient $\nabla f$ instead
of parallel transports or geodesic flows, i.e., approximations to the exponential map. Note also that there is freedom to choose a constant 
positive definite and symmetric metric $g$ that acts as
a \emph{preconditioner}, which can further improve convergence.
Algorithm~\ref{dissgeorattle} is of order $r=2$ and obeys
Corollary~\ref{thm_rate_preserv}.

\section{Numerical Example}
\label{sec:numerical}

\begin{figure}[t]\centering
\includegraphics[width=\textwidth,trim={0 10 10 12},clip]{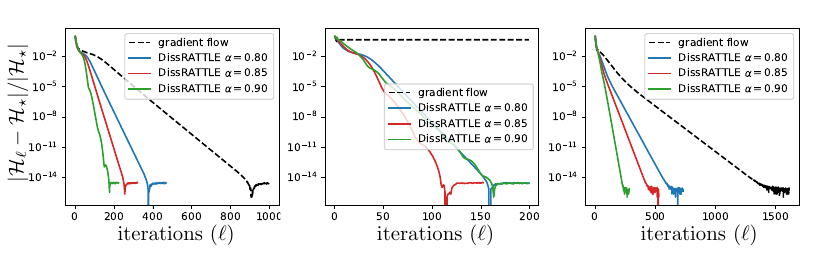}
\caption{\label{fig:ssk}
Solving problem \eqref{HSSK}/\eqref{ssk_prob} with Algorithm~\ref{dissgeorattle} and
Riemannian gradient descent.
We set $n=1000$ dimensions.
\emph{Left:} Step size $h=0.9/\lambda_{\text{max}}(M)$ for
both methods, and no external field ($\rho = 0$).
\emph{Middle:} Same problem but increased step size,
$h=1.9/\lambda_{\text{max}}(M)$.
\emph{Right:} We add an external field ($\rho=0.1$). We set
$h=0.9/\lambda_{\text{max}}(M)$ for both methods.
}
\end{figure}

We consider
an important model in the theory of disordered systems,
the spherical Sherrington-Kirkpatrick (SSK) model, which provides a baseline
for random optimization problems.  We follow the setup of
\cite{Baik:2021}. The system is described by a Hamiltonian (energy)
\be\label{HSSK}
\mathcal{H}(\bm{\sigma}) = -\dfrac{1}{2} \bm{\sigma}^T M \bm{\sigma} - \rho \, \bm{g}^T \bm{\sigma},
\ee
where $\bm{\sigma} \in \mathcal{S}_{n-1}$ lies on the hypersphere of radius
$\| \bm{\sigma} \| = \sqrt{n}$---note that $\bm{\sigma}$ plays the role of $\qq$ in our  previous notation. 
The disorder matrix $M$ is symmetric with i.i.d. entries $M_{ij} \sim 
\mathcal{N}(0, (1+\delta_{ij})/n)$, where $\delta_{ij}$ is the Kronecker delta.
The external field $\bm{g}$ has i.i.d. entries from a standard Gaussian, $\mathcal{N}(0,1)$.
The goal is to solve 
\be\label{ssk_prob}
\min_{\bm{\sigma} \in \mathcal{S}_{n-1}} \mathcal{H}(\bm{\sigma}) .
\ee
We compare Algorithm~\ref{dissgeorattle} with  Riemannian gradient descent
\cite{Sra:2016}, here denoted as ``gradient flow.''
In the absence of external field, $\rho=0$, it can be shown that 
the problem has an exact solution,
$\mathcal{H}_\star = -\tfrac{n}{2} \lambda_{\text{max}}(M)$.
When $\rho\ne0$ the problem does not have a closed form solution, thus
we set $\mathcal{H}_\star$ the smallest value obtained when $\ell \to \infty$.
In Fig.~\ref{fig:ssk} we show one instance of this problem.
For all methods, we choose the largest step size $h = 0.9/\lambda_{\text{max}}(M)$ allowed by gradient flow to converge (left and right plots). Algorithm~\ref{dissgeorattle}
is \emph{significantly faster and much more stable}, allowing even larger step sizes
(middle plot). We pick three different values of $\alpha$ for illustration.
The parameter $\alpha$ was not tuned at all, 
and if one carefully tune $h$ and $\alpha$,
Algorithm~\ref{dissgeorattle} is even faster than illustrated in these
plots. Details about the implementation as well
additional experiments are provided  in the Appendix.

\section{Conclusion}
\label{sec:conclusion}

We introduced a general framework for constructing optimization methods
over smooth manifolds and for problems with nonlinear (equality/inequality) constraints.
Our approach relies on a dissipative extension of Dirac's formalism of
constrained Hamiltonian systems. 
As a result, we derived a dissipative geodesic equation that is consistent with the
KKT conditions on Riemannian manifolds. Such a system was shown to have  
favorable stability and convergence rate properties. 

We also extended symplectic
integrators  to dissipative and constrained cases over 
manifolds. Such discretizations 
preserve the main properties of the continuum dynamics, and in particular
its convergence rates (up to a controlled error).
Based on these ideas, we derived a simple and efficient method based on a 
dissipative extension
of the famous RATTLE integrator. Such a method  is able to solve optimization 
problems  with equality and inequality constraints. Numerically, this 
method proved to be highly stable and  fast.

We deferred many technical details and additional results to 
the Appendix,  such as the proofs of the
main results, and a method for optimization over Lie groups.

Finally, this work lays out the foundations of a general approach
to constrained optimization on manifolds. 
There are many possible extensions and refinements of our results.  For instance, 
it would be interesting to obtain more general 
(global) convergence rate results for systems \eqref{dissip_geo} or
\eqref{dissip_geo2}, based on Theorem~\ref{thm_impliODE} in the Appendix.

\subsection*{Acknowledgements}
\vspace{-.5em}
This work is supported by the Army Research Office (ARO) under 
contract W911NF-17-1-0304 as
part of the collaboration between US DOD, UK MOD and UK Engineering and Physical
Research Council (EPSRC) under the Multidisciplinary University Research Initiative (MURI).

\bigskip


\appendix

\section{Review of Conservative Hamiltonian Systems} \label{sec:Cons}

In this section, we provide a brief review of the geometric formulation of
conservative Hamiltonian systems. The goal is to recall  basic concepts and introduce notation; we refer to \cite{Arnold:1989,Marsden:2010,Berndt:2000} for  details.
This paper  assumes familiarity with differential geometry and tensor 
manipulations as commonly found in general relativity \cite{Carroll:2019,Nakahara:2003}.

Given a smooth $n$-dimensional manifold $\M$, we can construct its
cotangent bundle $T^*\M$, which is the union of all cotangent spaces $T_{\mathfrak{p}}^* \M$
for all  points $\mathfrak{p} \in \M$.\footnote{The cotangent space 
$T_{\mathfrak{p}}^*\M$ is the  dual vector space to
the tangent space $T_{\mathfrak{p}}\M$.
It  exists independently of $T_{\mathfrak{p}}\M$ and cannot be identified with it 
without further structure, such as a Riemannian metric.}
The cotangent bundle 
 carries a canonical symplectic 
structure $\omega$ that will be defined shortly.
Around any point  $\mathfrak{p} \in \M$ there exists local coordinates $\qq=(q^1,\dotsc,q^n)$ on $\mathbb{R}^n$. We henceforth refer to $\mathfrak{p}$ simply by its 
coordinates $\qq$ when clear from the context.
The  cotangent bundle then inherits local coordinates $(\q^i, \p_i)$, for $i=1,\dotsc,n$,  around $T_{\qq}^*\M$ ($p_i$ denotes the momentum, which is the dual to
$q^i$). Thus, $T_{\qq}^*\M$ can be locally identified with 
$\mathbb{R}^{2n}$.
In such a coordinate basis we can write the symplectic 2-form\footnote{We use
Einstein's summation where repeated upper  and lower indices are summed
over, i.e.,  $u^i v_i \equiv \sum_i u^i v_i$. Also, upper indices, $u^i$,
denote components of vectors, $\bm{u} \in T\M$, whereas lower indices, $v_i$,
denote components of covectors (or dual vectors), $\bm{v} \in T^* \M$. }
\be \label{omega}
\omega \equiv  d\q^i \wedge d\p_i ,
\ee
where $\wedge$ denotes the exterior or wedge product between differential forms
\cite{Berndt:2000}.
This 2-form is nondegenerate since its rank is $2n$, and also closed, $d\omega = 0$ ($d$ is the exterior derivative), hence
 $(T^*\M, \omega)$ is a \emph{symplectic manifold} \cite{Berndt:2000}.
 This is always true, namely for any smooth manifold $\M$.
 Therefore, the cotangent bundle $T^*\M$ of any smooth manifold $\M$ is a symplectic manifold.

Given a function $H : T^*\M \to \mathbb{R}$, called Hamiltonian, the 2-form 
$\omega$ induces a dynamical system  obeying  Hamilton's equations,
\be \label{hamiltons}
\dot{\q}^i = \dfrac{\partial H}{\partial \p_i},  \qquad
\dot{\p}_i = -\dfrac{\partial  H}{\partial \q^i},
\ee
for $i=1,\dotsc,n$,
where $\dot\q^i \equiv d\q^i / dt$ and $t$ denotes  time parametrization.
It is immediate  that
\be \label{Hder}
\dfrac{d H}{dt} = 0,
\ee
i.e., $H$ is conserved.
Hamilton's equations \eqref{hamiltons} can be more concisely written as
\be \label{hamiltons_omega}
i_{X_H}(\omega) = - dH ,
\ee
where $i_{X_H}$ is the interior product and
$X_H$ is the Hamiltonian vector field, given by
\be \label{XH}
X_H \equiv \dot{\q}^i \dfrac{\partial}{\partial \q^i}
+ \dot{\p}_i  \dfrac{\partial }{\partial \p_i}.
\ee
 From \eqref{hamiltons_omega}, together with Cartan's formula, one
can easily show that the symplectic structure
is preserved, namely
$\Lie_{X_H} \omega = 0$, where $\Lie_{X_H}$ is the Lie derivative along 
the vector field $X_H$.
Conversely, one can
show that any vector field $X$ that preserves $\omega$ must, locally, obey
Eq.~\eqref{hamiltons_omega} for some function $H$, i.e.,
$X$ must be, at least locally, the vector field of a Hamiltonian
system ($X=X_H$).
This is the reason why Hamiltonian systems
are special: They are  the only dynamics that
preserve the canonical symplectic structure of the cotangent bundle $T^*\M$ of
any smooth manifold $\M$; this is essentially the proof of
Theorem~\ref{thm_local_ham}.

\subsection*{Symplectic integrators} This special class of numerical integrators 
consists of discretizations of conservative Hamiltonian systems required to 
exactly preserve the symplectic structure $\omega$ \cite{Leimkuhler:2004,Hairer:2010}.
Therefore, such discretizations can be seen as Hamiltonian systems themselves,
with a perturbed or \emph{shadow Hamiltonian} that is a formal
expansion around the true Hamiltonian of the system in terms
of the step size. 
In short, such a class of numerical integrators 
inherits all the benefits of being Hamiltonian.

\section{The Dissipative Constrained Geodesic Equation}
\label{sec:geo_der}

Here we derive the geodesic equation \eqref{dissip_geo}.
Consider the Hamiltonian \eqref{hamM} but with an additional mass term, i.e.,
\be
H_{\text{total}} = \dfrac{1}{2m} e^{-\eta} g^{ij} p_i p_j + e^{\eta} f + e^{\eta} \lambda^a \g_a.
\ee
Hamilton's equations yield
\begin{subequations} \label{hamm}
\begin{align}
\dot{q}^i &\approx \dfrac{\partial H_{\text{total}}}{\partial p_i} = 
\dfrac{1}{m} e^{-\eta} g^{ij} p_j, \label{hamm1}\\
\dot{p}_i &\approx -\dfrac{\partial H_{\text{total}}}{\partial q^i}
= - \dfrac{1}{2m} e^{-\eta} \partial_i g^{jk} p_j p_k
- e^{\eta} \partial_i f - e^{\eta} \lambda^a \partial_i \g_a . \label{hamm2}
\end{align}
\end{subequations}
These equations are subject to the constraints $\g_a(\qq) = 0$, hence the symbol
$\approx$ according to Dirac's notation.
Differentiating the first equation with respect to time,
\be
\ddot{q}^i \approx \dfrac{1}{m} e^{-\eta} g^{ij} \dot{p}_j
+ \dfrac{1}{m} e^{-\eta} \partial_k g^{ij} \dot{q}^k p_j 
- \dfrac{1}{m} \dot{\eta} e^{-\eta} g^{ij} p_j.
\ee
Replacing \eqref{hamm2},
\be
\ddot{q}^i \approx 
\dfrac{1}{m} e^{-\eta} \partial_k g^{ij} \dot{q}^k p_j 
 -\dfrac{1}{m} \dot{\eta} e^{-\eta} g^{ij} p_j
 -\dfrac{1}{2m^2} e^{-2\eta} g^{ij} \partial_j g^{k \ell} p_k p_\ell 
 - \dfrac{1}{m} g^{ij} \partial_j f - \dfrac{1}{m}\lambda^a  g^{ij} \partial_j \g_a  .
\ee
From \eqref{hamm1} we have
$p_i = m e^{\eta} g_{ij} \dot{q}^j$, which replaced above yields
\be
\ddot{q}^i - g_{j\ell} \partial_k g^{ij} \dot{q}^k  \dot{q}^\ell
+ \dfrac{1}{2}  g^{ij} \partial_j g_{k \ell}  
\dot{q}^k \dot{q}^\ell  +\dot{\eta} \dot{q}^i \approx 
- \dfrac{1}{m} g^{ij}\partial_j f - \dfrac{1}{m}\lambda^a g^{ij} \partial_j \g_a .
\ee
Using $g_{j\ell} \partial_k g^{ij} = - g^{ij} \partial_k g_{j\ell}$ and
the definition of Christoffel symbols,
\be
\Gamma_{ijk} \equiv \tfrac{1}{2}(
\partial_k g_{ij} + \partial_j g_{ik} - \partial_i g_{jk}), \qquad 
\Gamma\indices{^i_j_k} \equiv g^{im}\Gamma_{mjk},
\ee
this last equation can be written as
\be\label{ode_mass}
\ddot{q}^i + \Gamma\indices{^i_j_k} \dot{q}^j \dot{q}^k 
 +\dot{\eta} \dot{q}^i \approx - \dfrac{1}{m} g^{ij} \left( \partial_j f + \lambda^a  \partial_j \g_a \right).
\ee

\section{Shadow Property} 
\label{sec:shadow}

\subsection{Proof of Theorem~\ref{thm_ham_preserv}}
\begin{proof}
First,
consider the vector field of the symplectified system
\eqref{ham2},  
\be
\begin{split}
X_{\mathscr{}H} &= \dfrac{\partial \mathscr{H}}{\partial P_I}\dfrac{\partial}{\partial Q^I} - \dfrac{\partial \mathscr{H}}{\partial Q^I} \dfrac{\partial}{\partial P_I}\\
 &= 
\dfrac{\partial}{\partial \q^0}
- \dfrac{\partial H_{\text{total}}}{\partial \q^0} \dfrac{\partial}{ \partial \p_0} + X_H - 
\g_a \dfrac{\partial}{\partial \pi_a} \\
&\approx 
\dfrac{\partial}{\partial q^0} - \dfrac{\partial H}{\partial q^0} \dfrac{\partial}{\partial p_0} + X_H ,
\end{split}
\ee
where the last equality  
holds on the constraint surface $\g_a(t, \qq, \pp) = 0$ ($a=1,\dotsc,m$).
The flow of the first two terms yield
\be
\dfrac{d q^0}{d s} =1, \qquad 
\dfrac{d p_0}{d s} = - \dfrac{\partial H}{ \partial {q^0}} = - \dfrac{d H}{ds},
\ee
where we have made use of \eqref{noncons}.
Thus,
\be
q^0(s) = s, \qquad p_0(s) = - H(s) + H(0) + p_0(0)
\approx - H(s) + \mathscr H(0) ,
\ee
provided $q^0(0)=0$, which fixes  $q^0 = t = s$.
The Hamiltonian is determined up to an arbitrary constant, and
it is convenient to take $ \mathscr H (0)=0$ since then
$p_0$  simply corresponds to the value  $-H(s)$ at a given instant of time.
Thus, we have precisely the ``gauge fixing'' of Eq.~\eqref{gauge}.

A presymplectic integrator (see Definition~\ref{presymp_def}) is 
a \emph{restricted symplectic integrator} $\Phi_h$ applied to the higher-dimensional system
\eqref{ham2}.
Let $\widehat{\varphi}_{s}$ denote the true flow of $X_{{\mathscr H}}$.
Because $\Phi_h$  is symplectic, and moreover it is a variational integrator (see Assumption~\ref{assump_var}), 
it holds globally  that \cite{Benettin:1994,Marsden:2001}
\be  \label{flow1}
{\mathscr H} \circ \Phi^\ell_{h} = {\mathscr H} \circ \widehat{\varphi}_{h\ell} + \bigO(h^r)
\ee
for $s_\ell = \ell h = \bigO(h^r e^{c/h})$, $\ell=0,1,\dotsc$, and
some constant $c > 0$.
Recall that $q^0$ is just the time parameter, which is integrated
exactly, and $p_0$ is a function of time alone and it does not
couple to the other degrees of freedom, thus it is also 
 integrated exactly (this was noted in 
\cite{Marthinsen:2014,Asorey:1983} as well).
Denote 
\be\label{short}
\bm{Z} \equiv (\q^0, p_0, \q^i, p_i), \qquad
\bm{z} \equiv (q^i, p_i), \qquad
\bm{Z}_\bullet \equiv \bm{Z}(s=0).
\ee
We have
\be
q^0_\ell \equiv
(q^0 \circ \Phi_h^\ell )(\bm{Z}_\bullet)  =
q^0 \circ \widehat{\varphi}_{hl}(\bm{Z}_\bullet)
= h \ell,
\ee 
i.e., we just used the fact that $\Phi_h$ integrates $q^0$  exactly.
Similarly, 
\be \label{p0exact}
p_{0, \ell} \equiv (p_0 \circ \Phi_h^\ell)( \bm{Z}_\bullet) = p_0 \circ \widehat{\varphi}_{h\ell}(\bm{Z}_\bullet) = p_0(h \ell) ,
\ee
which is however irrelevant since this term does not couple to the other
degrees of freedom.
On the constraint surface 
$\mathscr{H}  \approx H(q^0, q^i, p_i)$,
which is a function on $\mathbb{R}\times T^*\Q $, the extended cotangent
bundle of the base manifold $\Q$.
Define  the projection 
\be\begin{split}
\pi :  \mathbb{R}^2 \times T^*\Q  & \to  \mathbb{R} \times  T^* \Q  \\
\bm{Z} & \mapsto (q^0, \bm{z}),
\end{split}
\ee 
which eliminates $p_0$. We can write
\be \label{Hpi}
{\mathscr{H}} \approx p_0 +  H \circ \pi
\ee
and
\begin{subequations}
\begin{align}
\pi \circ  \widehat{\varphi}_{h \ell} &=  \varphi_{h \ell} \circ \pi, 
\label{PhiPi1}
\\
\pi \circ \Phi_h^\ell &= \phi_h^\ell \circ \pi,
\label{PhiPi2}
\end{align}
\end{subequations}
where
$\varphi_{t}$ denotes the  true flow of the original dissipative Hamiltonian
system on $\mathbb{R} \times T^*\Q  $, and 
$\phi_h^\ell$ denotes $\ell$ iterations of the \emph{presymplectic} integrator
$\phi_h$ on $\mathbb{R} \times T^*\Q  $.
Hence,  from \eqref{flow1} and \eqref{Hpi}, 
\be
p_0\circ \Phi^\ell_{h} + H \circ \pi \circ \Phi^\ell_{h}  = 
p_0\circ \hat{\varphi}_{h  \ell} + H \circ \varphi_{h \ell} \circ \pi  + \bigO(h^r) ,
\ee
where we used \eqref{PhiPi1} on the 2nd term of the RHS.
Due to \eqref{p0exact} we obtain
\be
H \circ \pi \circ \Phi^\ell_{h}  =  H \circ \varphi_{h \ell} \circ \pi  + \bigO(h^r) ,
\ee
and upon using  \eqref{PhiPi2} on the LHS we finally have
\be \label{hamcons}
H \circ \phi^\ell_h  =  H \circ \varphi_{h \ell}   + \bigO(h^r) .
\ee
This proves the relation \eqref{ham_preserv}, which is thus a 
consequence of 
\eqref{flow1} (shadow) and the symplectification.
It should be noted, however, that this derivation assumes that all numerical maps 
and the Hamiltonian are globally
defined, which is implicit under Assumption~\ref{assump_var}.

The above argument shows that the value of the nonconserved Hamiltonian $H$ along the presymplectic integrator stays close to its true value.
Furthermore, from the shadow expansion 
\be
\widetilde{{\mathscr H}} = 
{\mathscr H} + h^{r} \Delta {\mathscr H}_r + \dotsm ,
\ee 
assumed to exist for the symplectic integrator $\Phi_h$ applied
to the conserved Hamiltonian ${\mathscr H}$ since it is a variational integrator,
upon using \eqref{Hpi} and \eqref{p0exact} we conclude that 
there exists a \emph{shadow nonconservative Hamiltonian} $\widetilde{H}$ given by
\be
\widetilde{H}   = H  +  h^r \Delta H_r + \dotsm  .
\ee
\end{proof}

The next result is a consequence of this shadow property and shows that 
presymplectic discretizations of system \eqref{dissip_geo} are able closely reproduce
its convergence rates, whatever they are.

\subsection{Proof of Corollary~\ref{thm_rate_preserv}}
\begin{proof}
Suppose the generic Hamiltonian system
\eqref{dissip_geo} has a  rate of convergence of the form
\be \label{cont_rate}
f(\qq(t)) - f(\qq_\star) = \bigO(R(t))
\ee
for some decreasing function $R(t) > 0$, which tells us how fast the system converges to a minimum
of the objective functions $f$.  One would like to preserve such a convergence rate
under discretization.
Let $\phi_h$ be a presymplectic integrator of
order $r \ge 1$ (Definition~\ref{presymp_def}). 
It then holds that \cite{Hansen:2011}
\be \label{dist}
\big\|  \phi_h^\ell(\bm{z}_\bullet) - \varphi_{t_\ell}(\bm{z}_\bullet) \big\| \le 
C_\ell h^r
\ee
for some constant $C_\ell > 0$ 
and sufficiently 
small step size $h > 0$,
where $\bm{z}_\bullet \equiv (\q^i(0), \p_i(0))$ is the initial state, 
$t_\ell = h \ell$ is the simulation time ($\ell=0,1,\dotsc$), and 
we recall that $\varphi_{t}$ is the true flow.
Note also that $\| \cdot \|$ in \eqref{dist} represents some norm defined
over the manifold $T^*\Q$ \cite{Hansen:2011}.
Replacing the Hamiltonian  \eqref{hamM}, i.e.,
$H = \tfrac{1}{2} e^{-\eta(t)} g^{ij} p_i p_j + e^{\eta(t) } f(q)$,
into Eq.~\eqref{ham_preserv}   yields
\be \label{rate_pf0}
 f \circ \phi_h^\ell - f \circ \varphi_{h\ell} =
 e^{-2 \eta(t_\ell) } \big( T\circ \varphi_{h\ell} - T \circ \phi_h^{\ell} \big) 
  + e^{- \eta(t_\ell) } h^r K  ,
\ee
for some constant $K > 0$ and
where $T \equiv \tfrac{1}{2} \pp \cdot g^{-1}(\qq) \pp$ is the kinetic energy.
Due to the smoothness of the Riemannian manifold $\Q$ and 
\eqref{dist} we have the Lipschitz condition
\be
\big| \big( T\circ \varphi_{h\ell} - T \circ \phi_h^{\ell} \big) \big|
\le L_T C_\ell h^r,
\ee
for some constant $L_T > 0$.
Therefore,
\be \label{rate_pf1}
\big| f \circ \phi_h^\ell - f \circ \varphi_{h\ell} \big| \le
 e^{- \eta(t_\ell) }  h^r \big( L_T C_\ell e^{-\eta(t_\ell)} + K \big) .
\ee
The result \eqref{preserv_rate} follows provided 
$C_\ell e^{-\eta(t_\ell)}$ is finite. Let us use a pretty conservative bound
on $C_\ell$, 
for instance
on $\mathbb{R}^n$ it is common to have 
$C_\ell = C (e^{ L_{\phi}  t_\ell } - 1)$,
where $L_{\phi}$ is a Lipschitz constant of the integrator \cite{Hairer:2010},
i.e., 
$\| \phi_h(z_1) - \phi_h(z_2) \| \le L_\phi \| z_1 - z_2\|$,
and $C > 0$
is a constant that does not depend on $\ell$.
Such a bound holds for the majority of methods, even nonstructure-preserving ones.  
Thus, under such a growth condition on $C_\ell$ and 
by the assumption in the theorem, i.e.,
$e^{L_\phi t_\ell - \eta(t_\ell)} < \infty$, we conclude that
\be \label{disc_rate}
\underbrace{R_\ell}_{\text{discrete}} = \underbrace{ R(h \ell) }_{\text{continuum}}  + \underbrace{ \bigO\big( e^{- \eta(t_\ell) } h^r\big) }_{\text{small error}} 
\ee
during suitable simulation times as stated in
Theorem~\ref{thm_ham_preserv}.
\end{proof}

\section{KKT Conditions and Stability}
\label{sec:KKT}

In this section, we describe the results related to the estability system \eqref{dissip_geo}, stated in Sec.~\ref{sec:diss_geo}.
Such results extends immediately to system \eqref{dissip_geo2} which incorporates
inequality constraints.
We start by characterizing critical points in terms of the KKT conditions.

\subsection{Proof of Theorem~\ref{thm_kkt1}}
\begin{proof}
The system \eqref{dissip_geo} can be written in first order form  as
\begin{subequations} \label{1storderform}
\begin{align}
\dot{\q}^i &= g^{ij}(\qq) \p_j, \\
\dot{\p}_i &= -\dot{\eta}(t) \p_i -\dfrac{1}{2} \partial_i g^{jk}(\qq) p_j p_k
- \partial_i f(\qq) - \lambda^a \partial_i  \g_a(\qq) , \\
0 &= \g_a(\qq) .
\end{align}
\end{subequations}
By definition,
a critical point $(\qq_\star, \lambdav_\star, \pp^\star)$ is 
such that the vector field on the RHS vanishes, namely
$\p^\star_i = 0$, $\g_a(\qq_\star) = 0$, and
$\partial_i f(\qq_\star) + \lambda_\star^a \partial_i \g_a(\qq_\star) = 0$,
for all $i=1,\dotsc,n$ and $a=1,\dotsc,m$.
The last two relations are precisely the KKT conditions 
for the optimization problem \eqref{opt_problem2}.
\end{proof}

This result shows that critical points of system \eqref{1storderform}
obey the KKT conditions, which are the first order \emph{necessary} conditions
for optimality.  Recall that if the system converges, 
it must do so to a critical point.

\subsection{Extension to inequality constraints}

Let us also show that the previous result extends to system
\eqref{dissip_geo2} with more general KKT conditions.

\begin{theorem}\label{thm_gen_kkt}
Let $(\qq_\star, \lambdav_\star, \bm{\mu}_\star)$ (and $\pp^\star = \bm{0}$) be a critical
point of system \eqref{dissip_geo2}. Then such a critical point
must obey the KKT conditions for the optimization problem
\be
\min_{\qq \in \M} f(\qq) \quad \textnormal{s.t.} \quad \gv(\qq) = \bm{0}, \qquad
\hhv(\qq) \le \bm{0},
\ee
namely
\begin{subequations}
\begin{align}
\partial_i f(\qq_\star) + \lambda_\star^a \partial_i \g_a(\qq_\star) 
+ \mu^b_\star \partial_i \hh_b(\qq_\star) &= 0 
\qquad \textnormal{(stationarity)}, \label{gkkt1} \\
\g_a(\qq_\star) = 0,  \quad \hh_b(\qq_\star) &\le  0 
\qquad \textnormal{(primal feasibility)}, \label{gkkt2} \\
\mu_\star^b \cdot \hh_b(\qq_\star) &= 0 
\qquad \textnormal{(complementary slackness)}, \label{gkkt3} \\
\mu_\star^b &\ge 0 \qquad \textnormal{(dual feasibility)}, \label{gkkt4}
\end{align}
\end{subequations}
where $i=1,\dotsc,n$, $a=1,\dotsc,m$, and $b=1,\dotsc,\overline{m}$.
\end{theorem}
\begin{proof}
We can write system \eqref{dissip_geo2} in first order as
\begin{subequations} \label{1storder_dissip_geo2}
\begin{align}
\dot{\q}^i &= g^{ij}(\qq) \p_j, \\
\dot{\p}_i &= -\dot{\eta}(t) \p_i -\dfrac{1}{2} \partial_i g^{jk}(\qq) p_j p_k
- \partial_i f(\qq) - \lambda^a \partial_i  \g_a(\qq) -\mu^b \partial_i \hh_b(\qq), \\
0 &= \g_a(\qq), \\
0 &= \hh_b(\qq),
\end{align}
\end{subequations}
under the condition that $\mu^b(t) = 0$ when $\phi_b(\qq(t)) < 0$, i.e., when
an inequality  constraint is \emph{inactive} it does not participate in the dynamics
and the system becomes unconstrained with respect to $\hh_b$.

A critical
point is a zero of the RHS of \eqref{1storder_dissip_geo2}, thus it must  obey $\p^\star_i = 0$, $\g_a(\qq_\star) = 0$, 
$ \hh_b(\qq_\star) = 0$, 
and  
$\partial_i f(\qq_\star) + \lambda^a_\star \partial_i \g_a(\qq_\star) + 
\mu^b_\star \partial_i \hh_b(\qq_\star) = 0$.
Note also that when $\mu^b_\star = 0$ we have $\hh_b(\qq_\star) < 0$, so 
the inequality $\hh_b(\qq_\star) \le 0$ is obeyed.
Hence, the KKT conditions \eqref{gkkt1} and \eqref{gkkt2} are satisfied.
Moreover, the dynamics is defined in such a way that
$\mu^b(t) \cdot \hh_b(\qq(t)) = 0$ at all times, and this holds
in particular at the critical point, thus condition \eqref{gkkt3} is obeyed.
Finally, as a consequence of this last equality, 
it was already shown in \eqref{Dab} and
\eqref{dual_feasibility} that  condition \eqref{gkkt4} follows.
\end{proof}

\subsection{Proof of Theorem~\ref{thm_stability}}

We now consider Lyapunov stability.
Note that it is implicit that $\dot\eta(t) > 0$ in order to have a dissipative
system---if $\dot{\eta}(t) < 0$ the system would be excited, i.e., energy would
be pumped into the system, and if $\dot{\eta}(t) = 0$
the system would be conservative. 

\begin{proof}
Consider  system \eqref{dissip_geo} written in first order form, i.e., Eq.~\eqref{1storderform} which we write as
\be
\dot{q}^i \approx g^{ij} p_j, \qquad \dot{p}_i \approx - \dot{\eta}(t) p_i - 
\dfrac{1}{2} \partial_i g^{jk} p_j p_k - \partial_i f - \lambda^a \partial_i \g_a 
\ee
following Dirac's notation.
Without loss of generality, assume the critical point of interest is at the origin, $\qq_\star = \bm{0}$, and that the minimum is achieved with value
$f(\qq_\star) = \min_{\qq \in \M} f(\qq) =  0$.
Furthermore, assume that the critical point is \emph{isolated}, namely there exists 
a neighborhood
of $(\pp^\star = \bm{0}, \qq_\star = \bm{0}, \lambdav_\star)$ such that the system
has no other critical point except for this one.
Now consider the function
\be
\mathcal{E}(\qq, \pp, \lambdav) \equiv \dfrac{1}{2} g^{ij}(\qq) p_i p_j + f(\qq) + \lambda^a \g_a(\qq) .
\ee
On the constraint surface, the last term is effectively zero, i.e., 
$\mathcal{E}(\qq_\star, \pp^\star, \lambdav_\star) \approx 0$, and also
\be
\mathcal{E}(\qq, \pp, \lambdav )  > 0
\ee 
for 
all points different than the critical point of interest. Thus $\mathcal{E}$ is
positive definite  in a sufficiently small region around the critical point.
Moreover, differentiating this function with respect to time and subbing the
equations of motion, we conclude that 
\be
\dot{\mathcal{E}} = - \dot{\eta}(t) g^{ij} p_i p_j \le 0.
\ee
The function $\mathcal{E}$ is therefore a Lyapunov function \cite{Wiggins:2000},
which implies that the  system is \emph{stable} on such a critical point. 
In addition, if $\eta(t) = \gamma t$ for $\gamma = \mbox{constant} > 0$, then LaSalle's
invariance principle \cite{Wiggins:2000} ensures that the system is
\emph{asymptotically stable} around the critical point, i.e.,
trajectories actually converge to the critical point.
\end{proof}

The exact same argument applies to system \eqref{dissip_geo2} on the more general 
setting of problem \eqref{opt_problem3}, which accounts
for inequality constraints. This is done by considering
the Lyapunov function
\be
\mathcal{E}(\qq, \pp, \lambdav, \bm{\mu}) \equiv \dfrac{1}{2} g^{ij}(\qq) p_i p_j + f(\qq) + \lambda^a \g_a(\qq) + \mu^b \hh_b(\qq).
\ee 
and making use of the equations of motion \eqref{1storder_dissip_geo2}.
The argument is unchanged since
the inequality constraint $\hh_b(\qq)$ works effectively as an equality
constraint, otherwise $\mu^b = 0$ and $\hh_b$ does not participate in the dynamics.  
Therefore, such a stability result remains valid in the presence of inequality constraints as well.

\section[Implicit Dynamics on the Constraint Surface and Convergence Rate]{Implicit Dynamics on the Constraint Surface \\ and Convergence Rate}
\label{sec:rate}

This section is perhaps the most technical, where we show the decay
rates of systems \eqref{dissip_geo} and \eqref{dissip_geo2} to localized
minima in the feasible region of the constraints.
The strategy is to solve explicitly for the Lagrange multipliers to obtain
a differential equation that describes the implicit dynamics on the constraint
surface.

We first consider the system \eqref{dissip_geo} and explicitly solve for the Lagrange multipliers, thus reducing it into a differential equation that characterizes the
dynamics on the constraint submanifold $\Q \subset \M$.
Denoting $\gamma(t) = \dot{\eta}(t)$, the constrained dissipative
geodesic equation of interest is
\be \label{dissgeo4}
\ddot{q}^i + \Gamma\indices{^i_j_k} \dot{q}^j \dot{q}^k + \gamma(t) \dot{q}^i = -\partial^i f - \lambda_a \partial^i \g^a, \qquad \g^a(\qq) = 0 .
\ee
We are now denoting $\g^a$ with upper indices since  we  view this as a
coordinate transformation. 
Define the ``vielbein'' fields \cite{Nakahara:2003,Carroll:2019}
\be
e\indices{^a_i} \equiv \dfrac{\partial \g^a}{\partial q^i} , \qquad 
e\indices{_a^i} \equiv \dfrac{\partial q^i}{\partial \g^a} .
\ee
They obey $e\indices{^a_i}e\indices{_a^j} = \delta^i_j$ and
$e\indices{_a^i} e\indices{^b_i} = \delta_a^b$.
The vielbein defines a  metric on the constraint surface (together with its inverse)
given by
\be \label{Gmetric}
G_{ab} \equiv e\indices{_a^i} g_{ij} e\indices{_b^j}, \qquad 
G^{ab} \equiv e\indices{^a_i} g^{ij} e\indices{^b_j}.
\ee
These objects  appear in the ``tetrad'' formalism of general relativity
\cite{Carroll:2019}.%
\footnote{In the same way that $g_{ij}$ lowers indices of the $\q^j$ coordinates, $G_{ab}$ lowers  indices of the $\g^b$ coordinates---and similarly $g^{ij}$ and $G^{ab}$ raises
indices of coordinates $\q_j$ and $\g_b$, respectively.
The expressions \eqref{Gmetric} can be seen as a change of metric,
$g \mapsto G$, in going from $\M$ to $\Q$.
}
The constraint $\g^a(\qq) = 0$ gives rise to the so-called
``hidden constraints'' $d \g^a / dt = 0$ and
$d^2 \g^a / dt^2 = 0$ that yield
\be
e\indices{^a_i} \dot{q}^i = 0
\ee
and 
\be \label{sec_hidden}
e\indices{^a_i} \ddot{q}^i + \partial_i e\indices{^a_j} \dot{q}^i \dot{q}^j  = 0 .
\ee 
Replacing the first of Eq.~\eqref{dissgeo4} above  allows
us to solve for 
the Lagrange multipliers,
\be\label{lambsol}
\lambda_c = G_{ca} \partial_i e\indices{^a_j}  \dot{q}^i \dot{q}^j - G_{ca} e\indices{^a_i} \left( \Gamma\indices{^i_j_k} \dot{q}^j \dot{q}^k + 
\gamma  \dot{q}^i + \partial^i f \right). 
\ee

Define the \emph{projection} operators
\be \label{PPdef}
\Ppar^{ij} \equiv g^{ij} - \Pper^{ij}, \qquad
\Pper^{ij} \equiv  e\indices{_a^i} G^{ab} e\indices{_b^j} .
\ee
One can check that they obey
$\Ppar^2 = \Ppar$, $\Pper^2 = \Pper$, and
$\Ppar \Pper = \Pper \Ppar = 0$.
The vielbeins  define the affine connection
\be\label{Gammaperpdef}
{\Gamma_{\perp}}\indices{^i_j_k} \equiv e\indices{_a^i} \partial_j e\indices{^a_k} .
\ee
Let us also define the projected connection
\be\label{Gammapardef}
{\Gamma_{\parallel}}\indices{^i_j_k} \equiv
\Ppar\indices{^i_\ell} \Gamma\indices{^\ell_j_k}. 
\ee
One can also check that
\be \label{PGamma}
\Pper{\Gamma_{\perp}} = {\Gamma_{\perp}}, \qquad
\Ppar \Gamma_{\perp} = 0, \qquad
\Pper \Gamma_{\parallel} = 0, \qquad
\Ppar \Gamma_{\parallel} = \Gamma_{\parallel},
\ee
and
\be \label{PGamma2}
{\Gamma_{\perp}}\Pper = {\Gamma_{\perp}}, \qquad
\Gamma_{\perp}\Ppar  = 0, \qquad
\Gamma_{\parallel}\Pper  = 0, \qquad
\Gamma_{\parallel}\Ppar  = \Gamma_{\parallel}.
\ee
Thus, $\Gamma_{\parallel}$ is a connection on the constraint surface that
defines the configuration manifold $\Q$, while $\Gamma_{\perp}$ is a connection
on its orthogonal complementary space.
With these ingredients, we  now obtain the following interesting result.

\begin{theorem}\label{thm_impliODE}
The  differential algebraic equation \eqref{dissgeo4} can be reduced to 
the differential equation
\be\label{impliODE}
\ddot{q}^i + \left( {\Gamma_{\parallel}}\indices{^i_j_k} + {\Gamma_{\perp}}\indices{^i_j_k}  \right) \dot{q}^j \dot{q}^k 
+ \gamma(t) \Ppar\indices{^i_j} \dot{q}^j = - \Ppar\indices{^i^j} \partial_j f,
\ee
which describes the implicit 
dynamics on the constraint subspace
$\Q$ (the configuration manifold) that is embedded into $\M$.
Furthermore, the dynamics is orthogonally decoupled into 
a driven dissipative geodesic equation and a free geodesic equation,
respectively  given by
\begin{subequations} \label{impliODE2}
\begin{align}
\Ppar\indices{^i_j}\ddot{q}^j +  {\Gamma_{\parallel}}\indices{^i_j_k}  \dot{q}^j \dot{q}^k 
+ \gamma(t) \Ppar\indices{^i_j} \dot{q}^j &= - \Ppar\indices{^i^j} \partial_j f , 
\label{impliODE21}\\
\Pper\indices{^i_j}\ddot{q}^j + {\Gamma_{\perp}}\indices{^i_j_k} \dot{q}^j \dot{q}^k
&= 0. \label{impliODE22}
\end{align}
\end{subequations}
\end{theorem}
\begin{proof}
Eq.~\eqref{impliODE}  follows by replacing the explicit form of the Lagrange multipliers
\eqref{lambsol} into \eqref{dissgeo4} and using
definitions \eqref{PPdef}--\eqref{Gammapardef}.
Eqs.~\eqref{impliODE2} follow from Eq.~\eqref{impliODE} by acting with
$\Ppar$ and $\Pper$ upon using the properties of these projection
operators and  relations \eqref{PGamma}.
\end{proof}

This result is completely general and may be of independent interest, e.g.,
it can provide the starting point to study the dynamics on more general
manifold and constrained optimization problems \eqref{opt_problem2}
and \eqref{opt_problem3}.
Note that \eqref{impliODE} and \eqref{impliODE2} no longer have mention to  Lagrange multipliers; 
these equations have  instead  projection operators to the constraint surface and
its orthogonal complement.
Moreover, Eq.~\eqref{impliODE22} is a free geodesic motion.
It implies that 
if the system is initialized on the constraint surface then it remains on
this surface at all times, i.e.,
if $\qq(0) = \Ppar \qq(0)$ and
$\dot{\qq}(0) = \Ppar \dot{\qq}(0)$ then 
$\Pper {\qq}(t) = 0$ and $\Pper\dot{\qq}(t) = 0$.
In this case, the dynamics is completely determined by Eq.~\eqref{impliODE21}
alone.
Based on this decomposition,
we can  finally study convergence to equilibrium using
stability theory since we have reduced the problem
to the analysis of a \emph{differential equation}.

\subsection{Proof of Theorem~\ref{thm_rate}}
\begin{proof}
By assumption, the initial state $\qq(0) \equiv \qq_0$ 
and $\dot{\qq}(0) = \bm{0}$ is on the 
constraint surface, therefore the dynamics is completely specified by Eq.~\eqref{impliODE21}. We have
\be \label{proj_states}
\qq = \Ppar \qq, \qquad 
\dot{\qq} \equiv \Ppar \dot{\qq}, \qquad 
\ddot{\qq} \equiv \Ppar \ddot{\qq},
\ee
since the state never leaves the constraint submanifold.  
Making use of the last equation in \eqref{PGamma2} we can  write
\be\label{odee}
\ddot{q}^i + {\Gamma_{\parallel}(\qq)}\indices{^i_j_k} 
\dot{q}^j
\dot{q}^k
+ \gamma \dot{q}^i = 
- \Ppar(\qq)\indices{^i^j} \partial_j f (\qq),
\ee
where here we consider $\gamma > 0$ constant.
We are interested in studying the system around an isolated critical point
$\qq_\star = \bm{0}$, assumed to be at the origin without loss of generality,
and which by definition obeys
\be
\Ppar(\qq_\star)\indices{^i^j} \partial_j f(\qq_\star) = 0.
\ee
A key result from stability theory is the 
Hartman-Grobmann theorem \cite{Hartman:1960},  allowing us to linearize
the system around $\qq_\star$ (the equilibrium is nondegenerate/hyperbolic by assumption).
Moreover, in a neighborhood of this point, there always exists 
Riemann normal coordinates
such that the connection vanishes, $\Gamma\indices{^i_j_k} = 0$ \cite{Nakahara:2003,Carroll:2019}.
This is the mathematical statement of the ``equivalent principle'' from
general relativity.
Additionally, the metric $g_{ij}$ can be taken to be the identity, $\delta_{ij}$, although for our purposes
we allow it to be a constant metric.
Using \eqref{proj_states} we thus have
\be
\ddot{q}^{i} + \gamma \dot{q}^i = - \Ppar(\qq)^{ij} \partial_j f(\Ppar \qq).
\ee
Expanding the RHS of this equation around $\qq_\star$ we have
the linearized system
\be\label{ode_lin}
\ddot{q}^i + \gamma \dot{q}^i = - \Ppar(\qq_\star)\indices{^i^j} 
\partial_j \partial_k f(\qq_\star) \Ppar(\qq_\star)\indices{^k_\ell} \q^\ell,
\ee
Note that $\Ppar$ defined in \eqref{PPdef} is symmetric,
therefore the operator on the RHS of \eqref{ode_lin} is (real) symmetric
and can be diagonalized by an orthogonal transformation.
Denote
\be \label{Hess_proj}
H_{\parallel} \equiv \big(\Ppar \nabla^2 f \Ppar\big)\big\vert_{\qq_\star}
\ee
the \emph{projected Hessian} at the critical point.
Letting $H_\parallel = O^T \Omega O$ be its diagonalization, and
defining $\bm{x} = O \qq$, the components of the 
differential equation \eqref{ode_lin} can decoupled and we are
left with several one-dimension problems of the form
\be \label{ode_one_dim}
\ddot{x} + \gamma \dot{x} = - \omega x, \qquad \omega \equiv 
\lambda(H_\parallel),
\ee
where $\omega$ denotes an eigenvalue of \eqref{Hess_proj}.
Around an isolated minimizer, all these eigenvalues are
positive.
Hence, the \emph{slowest}
degree of freedom of the system obeys this differential equation with
$\omega_{\text{min}}$, the smallest eigenvalue, which has solution
\be
x(t) = C_+ e^{\xi_+ t} + C_- e^{\xi_- t}, \qquad 
\xi = \dfrac{-\gamma \pm \sqrt{\gamma^2 - 4 \omega_\text{min}}}{2} ,
\ee
for constants $C_\pm$ depending on the initial conditions $\qq_0$ and $\dot{\qq}(0) = \bm{0}$.
Choosing $\gamma \le 2 \sqrt{\omega_{\text{min}}}$ (underdamped and critically damped regimes), from this exact solution we conclude that for all components
of the system we have
\be
\| \qq(t) - \qq_\star\| \le \| \qq_0 - \qq_\star \| 
e^{- \sqrt{ \omega_{\text{min}} } t } .
\ee
This show the result \eqref{rate} in terms of the smallest
eigenvalue of the projected Hessian \eqref{Hess_proj}. 

To make connection with the
operator \eqref{proj2}, consider explicitly the matrix
representation of $\Ppar$ in general.
From the definition \eqref{PPdef} we have
\be\label{Pmanip}
\begin{split}
\Ppar\indices{^i^j} &= g^{ij} - g^{ik} e\indices{^a_k} G_{ab} e\indices{^b_\ell} g^{\ell j} \\
&= g^{i k} \left( \delta^j_k - e\indices{^a_k} G_{a b} e\indices{^b_\ell} g^{\ell j}  \right) .
\end{split}
\ee
In our notation, $\Jac\indices{^a_i} = \partial \g^a/\partial q^i = e\indices{^a_i}$
is the $m \times n$ Jacobian matrix of constraints, $g^{ij}$ is the $n\times n$ inverse
metric $(g^{-1})_{ij}$, and $G_{ab}$ the $m\times m$ inverse matrix 
of $(G^{-1})^{ab}$.
From \eqref{Gmetric} we  have
\be
G^{-1} = \Jac g^{-1} \Jac^T \equiv \mathcal{R} .
\ee
We can thus write \eqref{Pmanip} in matrix form as
\be
\Ppar = g^{-1} \left(  I - \Jac^T \mathcal{R}^{-1} \Jac g^{-1}  \right)
= g^{-1} \mathcal{P}
\ee
in terms of the operator $\mathcal{P}$ defined in \eqref{proj2}.
Thus, the above eigenvalue can be alternatively represented by
\be \label{omegamin}
\omega_{\text{min}} \equiv 
\lambda_{\text{min}}\left( \left.\left(
g^{-1} \mathcal{P} \nabla^2 f \mathcal{P}^T g^{-1} 
\right) \right|_{\qq_\star} \right).
\ee
If we had fixed the metric to be Euclidean, $g_{ij} \vert_{\qq_\star} = \delta_{ij}$, 
then $\Ppar = \mathcal{P}$ at $\qq_\star$.
\end{proof}

Interestingly, the above result provides a local accelerated convergence
rate and it does not require  the Hessian of $f$ to be positive definite, i.e.,
it requires that the \emph{projected Hessian} in \eqref{omegamin} is positive definite,
which allows $f$ and the constraints to be \emph{nonconvex}.

In light of the previous discussions related to system
\eqref{dissip_geo2} that accounts for the inequality constraints
of problem \eqref{opt_problem3} (see, e.g., Eq.~\eqref{1storder_dissip_geo2}),
we know that inequality constraints are only  active when they are actually 
equality constraints. Therefore, the above result extends immediately to this case
as well, with the full operator \eqref{proj2} incorporating the nonlinear
constraints.

\section[Derivation of Dissipative RATTLE for Constrained Optimization]{Derivation of Dissipative RATTLE \\ for Constrained Optimization}
\label{sec:integrators}

Here we derive the algorithm introduced in Sec.~\ref{sec:diss_geo_rattle}.
We first consider the standard RATTLE integrator applied to conservative systems.
Then we reduce this method to our dissipative and constrained case.
We will then introduce a  modification of this method that is more efficient. 

\subsection{Dissipative RATTLE} \label{DisRattle}

Following Definition~\ref{presymp_def},
we first consider a symplectic integrator to conservative and
constrained systems 
based on the standard RATTLE method
\cite{Andersen:1983}; this method is known to be symplectic \cite{Leimkuhler:1994,Reich:1996} and it is also a variational
integrator (see Assumption~\ref{assump_var}).
This method has order of accuracy $r=2$.
Thus, consider a generic conservative and
constrained  system with total Hamiltonian
${\mathscr{H}}_{\text{total}} = 
\mathscr{H}(q^\mu, p_\mu) + \lambda^a {\g}_a(q^\mu)$, where $\mu=0,1,\dotsc,n$ and 
the constraints are holonomic.
We use the general form of this method written in \cite{Hairer:2010} 
that  reads
\begin{subequations} \label{rattle1}
\begin{align}
\p_{\mu, \ell+1/2} &= 
\p_{\mu, \ell} - \dfrac{h}{2}\left[ \dfrac{ \partial \Hc}{\partial \q^\mu}(\qq_\ell, \pp_{\ell+1/2}) +  \lambda^a_\ell \dfrac{\partial {\g}_a}{\partial \q^\mu}(\qq_\ell)  \right], \\
\q^{\mu}_{\ell+1} &= \q^\mu_{\ell} + \dfrac{h}{2}\left[
\dfrac{\partial \Hc}{\partial \p_\mu}(\qq_\ell, \pp_{\ell+1/2})
+\dfrac{\partial \Hc}{\partial \p_\mu}(\qq_{\ell+1}, \pp_{\ell+1/2}) \right], \\
0 &= {\g}_a(\qq_{\ell+1}) \qquad (a=1,\dotsc,m),
\label{c1} \\
\p_{\mu, \ell+1} &= \p_{\mu, \ell+1/2} - \dfrac{h}{2}\left[ \dfrac{ \partial \Hc}{\partial \q^\mu}(\qq_{\ell+1}, \pp_{\ell+1/2}) +  \rho^a_\ell \dfrac{\partial {\g}_a}{\partial \q^\mu}(\qq_{\ell+1})  \right], \\
0 &= \dfrac{\partial {\g}_a}{\partial \q^\mu}(\qq_{\ell+1}) \dfrac{\partial \Hc}{\partial p_\mu}(\qq_{\ell+1},\pp_{\ell+1}) \qquad (a=1,\dotsc,m),
\label{c2}
\end{align}
\end{subequations}
where $\ell=0,1,\dotsc$ is the iteration number.
Note that we introduced additional Lagrange multipliers $\rho^a$
associated to the so-called ``hidden constraints'' 
\be
0 = \dfrac{d {\psi}_a}{ds}  =  
\dfrac{\partial {\psi}_a}{\partial q^\mu} \dfrac{d q^\mu}{ds} = 
\dfrac{\partial {\psi}_a}{\partial q^\mu} \dfrac{d \Hc}{d p_\mu} ,
\ee
which forces the velocity  $dq^\mu / ds$ to be tangent to the
constraint surface.\footnote{We 
note that  we are implicitly assuming that the
Lagrange multipliers can be uniquely determined, which
is guaranteed when
 $\partial_{\qq} \gv ( \partial_{\pp\pp} \Hc )  \partial_{\qq} \gv^T$ is
 invertible. This condition is standard and also assumed for all
 constrained symplectic integrators in the literature.}
 This condition is not strictly necessary, but it results in a more stable
 integrator.

According to Definition~\ref{presymp_def}, 
to obtain a method for system \eqref{hamM},  all we have to do is set
$q^0_\ell = t_\ell = s_\ell = h\ell$ into \eqref{rattle1} and forget about
$p_{0}$ since it does not couple to the other degrees of freedom;
recall the gauge fixing \eqref{gauge}.
Moreover, to avoid dealing with implicit schemes in this paper, we will
assume that the metric $g_{ij}(\qq)$ is constant, i.e., independent of $\qq$,
so that all updates become explicit.
We thus obtain
\begin{subequations} \label{dissip_geo_rattle}
\begin{align}
\p_{i, \ell+1/2} &= \p_{i, \ell} - \dfrac{h}{2} e^{\eta( t_\ell)}
\left[
 \dfrac{\partial f (\qq_\ell)}{\partial q^i}
+ \lambda_\ell^a \dfrac{\partial \g_a(\qq_\ell)}{\partial q^i}
\right], \label{r1} \\
t_{\ell+1} &= t_\ell + h, \label{r2} \\
\q^{i}_{\ell+1} &= \q^i_{\ell} + \dfrac{h}{2}\left[
e^{-\eta(t_\ell) }  +
e^{- \eta( t_{\ell+1}) }   \right] g^{ij} p_{j,\ell+1/2},  \label{r3} \\
0 &= \g_a(\qq_{\ell+1}) \qquad (a=1,\dotsc,m), \label{r4} \\
\p_{i, \ell+1} &= \p_{i, \ell+1/2} - \dfrac{h}{2} e^{ \eta(  t_{\ell+1} )}\left[
\dfrac{\partial f (\qq_{\ell+1})}{\partial q^i} +
\rho^a_\ell \dfrac{\partial \g_a(\qq_{\ell+1})}{\partial \q^i}
\right],
\label{r5} \\
0 &= \dfrac{\partial \g_a(\qq_{\ell+1})}{\partial \q^i}  g^{ij} p_{j,\ell+1}
 \qquad (a=1,\dotsc,m).
 \label{r6}
\end{align}
\end{subequations}
Redefining the momentum variable as
\be \label{momchange}
e^{-\eta(t_\ell)} \pp_\ell \mapsto \pp_\ell,
\ee
and also defining the functions
\be \label{alpha}
\alpha_{\ell+1/2} \equiv e^{- \left(\eta(t_{\ell+1/2}) - \eta(t_{\ell}) \right)}
\ee
and
\be \label{beta}
\beta_{\ell+1} \equiv \dfrac{ e^{\eta(t_{\ell+1/2})-\eta(t_\ell)} + e^{-\left( \eta(t_{\ell+1}) - \eta(t_{\ell+1/2})\right)} }{2}
= \dfrac{(\alpha_{\ell+1/2})^{-1} + \alpha_{\ell+1}}{2},
\ee
we can write \eqref{dissip_geo_rattle} as
\begin{subequations} \label{algoDRat0}
\begin{align}
\pp_{\ell+1/2} &= \alpha_{\ell+1/2} \left[ \pp_{\ell} - (h/2) \nabla f(\qq_\ell) -
(h/2) \Jac(\qq_\ell)^T \bm{\lambda}_\ell \right], \label{algoDRat01} \\
\qq_{\ell+1} &= \qq_{\ell} + h \beta_{\ell+1} g^{-1} \pp_{\ell+1/2}, \label{algoDRat02} \\
\bm{0} &= \gv(\qq_{\ell+1}), \label{algoDRat03} \\
\pp_{\ell+1} &= \mathcal{P}(\qq_{\ell+1})\left[ \alpha_{\ell+1} \pp_{\ell+1/2} -
(h/2) \nabla f(\qq_{\ell+1}) \right] , \label{algoDRat04}.
\end{align}
\end{subequations}
%
%
%
Note that we wrote the method in vectorial form, and
some points need clarification.
For instance,  $\mathcal{J}(\qq)$ denotes the Jacobian matrix of constraints,
$\mathcal{J}_{ai}(\qq)  \equiv \partial \psi_a / \partial q^i\big\vert_{\qq}$,
and we introduced the projection operator
\be \label{projection}
\mathcal{P}(\qq) \equiv I - \mathcal{J}(\qq)^{T}\mathcal{R}(\qq)^{-1} \mathcal{J}(\qq) g^{-1} ,
\qquad
\mathcal{R}(\qq) \equiv \mathcal{J}(\qq) g^{-1} \mathcal{J}(\qq)^T .
\ee
This operator arises when solving \eqref{r6} explicitly, i.e.,  replacing
\eqref{r5} into \eqref{r6} and solving for the Lagrange multipliers $\bm{\rho}$.
The Lagrange multipliers $\bm{\lambda}$ are obtained by solving
\eqref{algoDRat03} numerically, i.e., the first three updates are solved
simultaneously. 
Note also that we left the function $\eta(t)$ completely arbitrary.

The above method proved to be very efficient in our experiments. However, we use it
as a stepping stone to construct an even more interesting method below, with
the same computational cost.

\subsection{Dissipative geodesic  RATTLE} \label{DisGeoRattle}

It is  interesting to consider a  more  modern 
formulations of RATTLE as proposed in molecular dynamics
\cite{Leimkuhler:2016}. 
When the Hamiltonian is separable,
it is convenient to  split the  potential and kinetic contributions since the flow of the former can be integrated exactly,  while the flow of the latter corresponds
to a free geodesic motion on the constraint surface. Thus,  consider
the constrained Hamiltonian
\be 
\label{Hbargeo}
{\Hc}  = \dfrac{1}{2} e^{- \eta( q^0 ) }
g^{ij} p_i p_j  + e^{\eta ( q^0 ) } f(q^i)
+ e^{\eta( q^0 )} \lambda^a  \g_a(q^i) + p_0 ,
\ee
which we will split with the sub-Hamiltonians
\begin{subequations} 
\begin{align}
\label{Hbarsplit}
{\Hc} &= {\Hc}_1 + {\Hc}_2, \\
\label{H12_def}
{\Hc}_1 &= e^{\eta( q^0 )} f(q^i) +  e^{\eta ( q^0 )} \nu^a \g_a(q^i), \\
\label{H22_def}
{\Hc}_2 &= \dfrac{1}{2} e^{-\eta( \q^0 )} g^{ij} p_i p_j
+  e^{\eta (  q^0 )} \lambda^a \g_a(q^i) + p_0 ,
\end{align}
\end{subequations}
where we introduced new multipliers $\nu^a$ in ${\Hc}_1$---the value
of the Lagrange multipliers are irrelevant when the constraints are satisfied.  Consider the  composition
\be \label{compos}
e^{h \Lie_{{\Hc}}} =
e^{( h/2) \Lie_{{\Hc}_1}}
e^{ h \Lie_{{\Hc}_2}}
e^{( h/2) \Lie_{{\Hc}_1}} + \bigO(h^3),
\ee
where $\Lie_{{\Hc}}$ denotes the Lie derivative along the flow  of ${\Hc}$. As will be clear
shortly, the flow of ${\Hc}_1$ can be integrated exactly, and we can replace any second order integrator
to approximate the flow of ${\Hc}_2$ without spoiling the
error in this approximation; we will use the dissipative
version of RATTLE obtained in \eqref{dissip_geo_rattle}.

The equations of motion related to \eqref{H12_def} are
\be \label{f1}
\dfrac{d \q^i}{ds} = 0, \qquad
\dfrac{ d \q^0}{ds} = 0, \qquad
\dfrac{ d\p_i}{ds} = e^{\eta( q^0 )}
\left( \dfrac{\partial f}{\partial q^i} + \nu^a
\dfrac{\partial \g_a}{\partial q^i} \right),  \qquad
 \g_a(\qq) = 0 .
\ee
In these equations,  only the momentum $\p_i$ evolves,  thus
the last equation is redundant
if the initial position already satisfies
the constraints. The original system
\eqref{Hbargeo} also  satisfies the hidden constraints
$d \psi_a / ds = 0$, i.e.,
\be \label{hidd}
\dfrac{\partial \psi_a}{\partial q^{i}} g^{ij} p_j = 0.
\ee
Upon differentiating this equation with respect to time  we can solve
for the Lagrange multipliers $\nu^a$ explicitly, yielding
\be \label{eq_flow1}
\dfrac{d \p_i}{ds} = - e^{\eta( q^0 )} \tensor{\mathcal{P}}{_i^j} 
 \dfrac{\partial  f}{\partial q^j},
\ee
where we have used the projection operator defined
in Eq.~\eqref{projection}. Since $\q^\mu$ is constant
we can integrate this equation exactly during a  time interval $\Delta s$,
\be \label{exact_sol_flow1}
\p_i(s+\Delta s) = \p_i(s)  -  (\Delta s) e^{ \eta( q^0(s) ) } 
\tensor{\mathcal{P}}{_i^j} (\qq(s))  \dfrac{\partial  f(\qq(s))}{\partial q^j} .
\ee

The equations of motion related  to  \eqref{H22_def} are
\be \label{f2}
\dfrac{d\q^i}{ds} = e^{- \eta(  q^0 ) } g^{ij} \p_j, \qquad
\dfrac{d\q^0}{ds} = 1, \qquad
\dfrac{d\p_i}{ds} = - e^{\eta ( q^0 ) }  \lambda^a \dfrac{\partial \g_a}{\partial q^i}, \qquad
\psi_a(\qq)  = 0.
\ee
When we set $q^0=s=t$ this corresponds to a  free dissipative motion
on the constraint surface,   i.e., without 
an external potential.  We can  numerically solve these
equations with
the dissipative RATTLE proposed in \eqref{dissip_geo_rattle} by setting   $f=0$.
Thus, combining the exact solution \eqref{exact_sol_flow1}
with the method \eqref{dissip_geo_rattle}  in approximating the composition
\eqref{compos},  within the same $\bigO(h^3)$ local error,
we  obtain
\begin{subequations}
\begin{align}
\pp_{\ell + 1/2} &= \pp_\ell - (h/2) e^{\eta(t_\ell)} \mathcal{P}(\qq_\ell) \nabla f(\qq_\ell), \\
\widetilde{\pp}_{\ell+1/2} &= \pp_{\ell+1/2} - (h/2) e^{\eta( t_\ell )}  
\mathcal{J}(\qq_\ell)^T \bm{\lambda}_\ell, \\
t_{\ell+1} &= t_\ell + h, \\
\qq_{\ell+1} &= \qq_{\ell} + (h/2) \big[ e^{-\eta( t_\ell ) } + e^{-\eta( t_{\ell+1} ) }  \big] g^{-1}  \widetilde{\pp}_{\ell+1/2}, \\
\bm{0} &= \gv(\qq_{\ell+1}), \\
\widetilde{\pp}_{\ell+1} &= \widetilde{\pp}_{\ell+1/2} - (h/2) \mathcal{J}(\qq_{\ell+1})^T \bm{\rho}_{\ell}, \label{geo6} \\
\bm{0} &= \mathcal{J}(\qq_{\ell+1}) g^{-1} \widetilde{\pp}_{\ell+1}, \label{geo7} \\
\pp_{\ell+1} &= \widetilde{\pp}_{\ell+1} - (h/2) e^{-\eta( t_{\ell+1} ) } \mathcal{P}(\qq_{\ell+1}) \nabla f(\qq_{\ell+1}) , \label{geo8}
\end{align}
\end{subequations}
where $\mathcal{J}$ is the Jacobian of constraints
and we used the projection operator \eqref{projection}.
We can further replace \eqref{geo6} into \eqref{geo7} and solve
for $\bm{\rho}_\ell$ explicitly, resulting  into
$\widetilde{\pp}_{\ell+1} = \mathcal{P}(\qq_{\ell+1}) \widetilde{\pp}_{\ell+1/2}$, which can now be directly combined with \eqref{geo8} to obtain
\be
\pp_{\ell+1} = \mathcal{P}(\qq_{\ell+1})\big[ \widetilde{\pp}_{\ell+1/2} - (h/2) e^{-\eta( t_{\ell+1} ) } \nabla f(\qq_{\ell+1}) \big].
\ee
Using the functions \eqref{alpha} and \eqref{beta},  and introducing the change of momentum variable 
\eqref{momchange},
we finally obtain
\begin{subequations} \label{geodesicRATTLE}
\begin{align}
\pp_{\ell+1/2} &= \alpha_{\ell+1/2}  \mathcal{P}(\qq_{\ell})\big[ \pp_{\ell}  -
(h/2) \nabla f(\qq_{\ell})\big], \label{algo1} \\
\widetilde \pp_{\ell+1/2} &=  \pp_{\ell+1/2}-(h \alpha_{\ell+1/2} /2) \mathcal{J}(\qq_\ell)^T \bm{\lambda}_\ell , \label{algo2} \\
\qq_{\ell+1} &= \qq_{\ell} + h \beta_{\ell+1} g^{-1} \widetilde{\pp}_{\ell+1/2}, \label{algo3} \\
\bm{0} &= \gv(\qq_{\ell+1}), \label{algo4} \\
\pp_{\ell+1} &= \mathcal{P}(\qq_{\ell+1})\big[ \alpha_{\ell+1} \widetilde{\pp}_{\ell+1/2} -
(h/2) \nabla f(\qq_{\ell+1}) \big] \label{algo5}.
\end{align}
\end{subequations}
The main benefit of this formulation compared to
\eqref{algoDRat0} is that the nonlinear equation
\eqref{algo4} is more easily satisfied since the projection in $\eqref{algo1}$ ensures that the initial momentum $\pp_{\ell +1/2}$ lies in the cotangent bundle of the manifold, and thus helps 
ensuring that $\qq_{\ell+1}$ stays close to the manifold
(we  verified this feature numerically).
Moreover, as previously mentioned, the component related to the potential
$\nabla f(\qq)$ is integrated exactly in \eqref{algo1} and \eqref{algo5}---%
recall \eqref{eq_flow1} and \eqref{exact_sol_flow1}.
For optimization purposes, we restore the mass $m=h$---see Eq.~\eqref{hamm} and
Eqs.~\eqref{hC2}--\eqref{lower_bound}---which only drops the step size in update
\eqref{algo3}. Note also that here $\alpha$ and $\beta$ are arbitrary,
depending on the dissipation function $\eta(t)$ which is generic---see
Eqs.~\eqref{alpha} and \eqref{beta}. Recalling \eqref{Jactot}, we thus have
Algorithm~\ref{dissgeorattle2}.

\begin{algorithm}[t]
\caption{\label{dissgeorattle2}
\textsc{DissRATTLE} is a presymplectic integrator for 
 problem \eqref{opt_problem4}. The algorithm accepts a sequence
 of functions $\{ \alpha_{\ell} \}$ where 
$\alpha_\ell \in (0,1)$ (momentum factor) and step size $h > 0$.
The parameter $\beta_\ell$ is fixed according to \eqref{beta}, and $g\succ 0$ is an
arbitrary  symmetric matrix (preconditioner).}
\begin{algorithmic}[1]
\For{$\ell=0,1,\dotsc$}
\State $\pp_{\ell+1/2} \leftarrow \alpha_{\ell+1/2}  \mathcal{P}(\qq_{\ell})\big[ \pp_{\ell}  -
(h/2) \nabla f(\qq_{\ell})\big]$ \label{geo11}
\State $\widetilde{\pp}_{\ell+1/2} \leftarrow  \pp_{\ell+1/2}-(h \alpha_{\ell+1/2} /2) \mathcal{J}(\qq_\ell)^T \bm{\Lambda}_\ell$  \label{geo12} 
\State $\qq_{\ell+1} \leftarrow \qq_{\ell} + \beta_{\ell+1} g^{-1} \widetilde{\pp}_{\ell+1/2}$  \label{geo13}
\For{$a=1,\dotsc,m+\bar{m}$}
\If{$\Psi_a$ is active}
\State $\Lambda_{a,\ell} \leftarrow \Psi_a(\qq_{\ell+1}) = 0$
\Else
\State $\Lambda_{a,\ell} \leftarrow 0$
\EndIf
\EndFor
\State $\pp_{\ell+1} \leftarrow \mathcal{P}(\qq_{\ell+1})\big[ \alpha_{\ell+1} \widetilde{\pp}_{\ell+1/2} -
(h/2) \nabla f(\qq_{\ell+1}) \big]$ \label{geo15}
\EndFor
\end{algorithmic}
\end{algorithm}

In Algorithm~\ref{dissgeorattle} we explicitly set
$\eta(t) = \gamma t$, with $\gamma = \mbox{const.} > 0$, which  yields
\be\label{alphabetaconst}
\alpha = e^{- \gamma h / 2}, \qquad \beta = \cosh(\gamma h / 2). 
\ee
Instead of tuning $\gamma$ and $h$,
the method tunes $\alpha$ and $h$, and then $\beta$ is fixed 
as $\beta = \cosh(- \log \alpha)$.  
Finally, the method \eqref{geodesicRATTLE} incorporates equality constraints
$\g_a(\qq) = 0$. 
However, the discussion in Sec.~\ref{sec:gen_kkt} shows that inequality
constraints, $\hh_b(\qq) \le 0$, work effectively as equality constraints or
do not play a role at all (the associated Lagrange multiplier is set to zero).
Accounting for this observation led us to redefine the Jacobian of constraints
as in Eq.~\eqref{Jactot}, and also the ``for'' loop between lines 5--11
in Algorithm~\ref{dissgeorattle2}.

\subsection{Numerical stability}

We provide a linear stability analysis
of the integrator 
\eqref{geodesicRATTLE} without constraints for simplicity since
the lower bound \eqref{lower_bound} is only known in this
setting.\footnote{The calculation can be done with constraints, but it becomes long
and  tedius, requiring solving for the Lagrange multipliers.} This provides an estimate
of the constant $C$ in the step size choice \eqref{hC2}.
Thus, it is sufficient to consider a
one-dimensional function $f = \tfrac{\omega}{2} \q^2$. 
We have
\begin{subequations}
\begin{align}
\p_{\ell+1/2} &= \alpha \left(\p_\ell - (h\omega/2) \q_\ell\right), \\
\q_{\ell+1} &= \q_\ell + h \beta m^{-1} \p_{\ell+1/2}, \\
\p_{\ell+1} &= \alpha \p_{\ell+1/2} - (h \omega/2) \q_{\ell+1}.
\end{align}
\end{subequations}
Replacing the first update into the subsequent ones we have
\begin{subequations}
\begin{align}
\q_{\ell+1} &= \left( 1 - h^2 \beta \alpha m^{-1} \omega/2 \right) \q_\ell + h\beta \alpha m^{-1} \p_\ell , \label{stab1} \\
\p_{\ell+1} &= \alpha^2 \p_\ell - (h\omega / 2) (1 + \alpha^2) \q_{\ell+1} .
\end{align}
\end{subequations}
For $\alpha \in (0,1)$ we have $\alpha \beta < 1$, hence  stability
requires a bound on the first term in update \eqref{stab1}, 
\be
\left| 1 - \dfrac{ h^2 m^{-1} \omega }{2} \right| \le 1 ,
\ee
hence
\be
 \dfrac{h^2}{m} \le \dfrac{4}{ \omega} ,
\ee
allowing us to choose the step size in \eqref{hC2} with a 
constant $C \le 2$.
We mention that other integrators, e.g., one based on the
first order symplectic Euler method would instead 
give $C \le \sqrt{2}$. It is interesting that
our method, which is based on an integrator of order
two and can be traced back to the leapfrog method, yields precisely the maximum step size able to match the lower bound \eqref{lower_bound}.

\section{Optimization over Lie Groups}
\label{sec:lie_group}

The reason for transforming the optimization problem \eqref{opt_problem}
over the configuration manifold $\Q$ into the constrained optimization
problem \eqref{constraints} over $\mathbb{R}^n$ is convenience: 
In general, we want to avoid 
computing geodesic flows, affine connections, parallel transports, etc., which
are numerically prohibitive, specially in higher dimensions.
However, there is a particular case of interest 
where the geodesic flow can be computed efficiently,
namely when the configuration manifold is a Lie group, $\Q \equiv \mathcal{G}$. 

Many Lie groups have Riemannian metrics that   are simple enough to allow tractable computations of parallel
transports, while being nontrivial  on a global topological level.
This is because its tangent space at  identity, i.e., its Lie
algebra $\liealg$, can be transported around the manifold by the group action,
and for Riemannian with sufficiently many 
symmetries  we can find tractable solutions for the geodesic flow on $\liealg$, via the Euler--Arnold equation, and transport the solutions over the manifold.

In this case we can consider an implicit dynamics directly on $\mathcal{G}$, rather than on some ambient space with
  constraints.
We  thus introduce a dissipative flow over a Lie group described by the time-dependent Hamiltonian\footnote{The $1/4$ factor in the kinetic energy
is because we do not require the basis of the Lie algebra to be normalized to have unit norm.}
\be \label{ham_lie}
H = -\dfrac{1}{4 }  e^{-\eta( t )} \Tr\big( V^T g^{-1} V\big) +  e^{\eta (t)} f(X)
\ee
with  $g$ being a symmetric positive definite constant matrix,
 $X \in \group$, and $V \in \liealg$ plays the role of the velocity/momentum and 
 belongs to the Lie algebra. 
 Note that $X$ and $V$ are (matrix) dynamical variables
that evolve in time.
The equations of motion over such a matrix  Lie group are 
\be \label{lie_mov}
\dot{X} =  e^{-\eta( t )}   X g^{-1} V, \qquad
\dot{V}  = - e^{\eta( t )} \Tr( \partial_X f(X)   X  T_a) T_a,
\ee
where $\{ T_a \}$ are the generators of the Lie algebra, obeying 
$[T_a, T_b] = i \tensor{C}{_a_b^c} T_c$ and assumed
to form an orthogonal basis, $\Tr(T_a T_b) \propto \delta_{ab}$.
For a practical implementation we actually do not need to use the
generators but rather a projection to the Lie algebra, 
\be \label{trace_form}
\Tr\left( \partial_X f(X)  X  T_a  \right)  T_a =
\left( \partial_X f(X)  X \right)^T - \partial_{X} f(X)  X .
\ee
Note also that   $\partial_X f$ denotes  a matrix with entries
$(\partial_X f)_{ij} = \partial f / \partial X^{ij}$.

\subsection{Dissipative Leapfrog over Lie Groups}
Similarly to the derivations in the previous section, we can construct a \emph{presymplectic}
integrator for system \eqref{lie_mov} by relying on a symplectic integrator.
Consider the dissipative version of RATTLE  given by 
 \eqref{dissip_geo_rattle} but without constraints. In this case
 the method reduces to the a dissipative version of the leapfrog.
We will proceed by analogy to simplify the discussion, although it is not
hard to formally justify the following steps.
Due to the form of the equations of motion  \eqref{lie_mov}
let us make the following correspondence:
\begin{subequations}
\begin{align}
\qq &\mapsto X,  \\
\pp &\mapsto Y, \\
\nabla  f(\qq)  &\mapsto \Tr\left( \partial_X f(X)  X  T_a  \right) T_a .
\end{align}
\end{subequations}
The update \eqref{r3}  can be seen as
\be 
\qq(t + h) = \qq(t) + (h/2) 
\big( e^{-\eta( t )} + e^{-\eta(t+h)} \big) g^{-1} \pp(t+h/2) 
\ee
and its leading order Lie group analog based on the equations of motion \eqref{lie_mov} reads
\be
X(t+h) = X(t) + 
\dfrac{h}{2} \big( e^{-\eta( t )} + e^{-\eta(t+h)}  \big)    X(t) g^{-1} V(t+h/2)
+ \dotsm .
\ee
However, this is only the first order expansion of the correct equation
obtained by actually exponentiating the Lie algebra, 
\be
X(t+h) = X(t) \exp\left\{
\dfrac{ h }{2} \big( e^{-\eta(t)} + e^{-\eta(t+h)}  \big) g^{-1}    V(t+h/2)
\right\}.
\ee
Therefore, with these  identifications, we obtain
the (unconstrained) Lie group version of the method \eqref{dissip_geo_rattle}
given by
\begin{subequations} \label{lie_leap1}
\begin{align}
V_{\ell + 1/2} &= V_{\ell} - (h/2) e^{\eta( t_\ell )} 
\Tr\left( \partial_X f(X_\ell)  X_\ell  T_a  \right) T_a, \\
t_{\ell + 1} &= t_\ell + h, \\
X_{\ell+1} &= X_\ell \exp\left\{ (h /2) \big(e^{-\eta( t_{\ell} )} + e^{-\eta(  t_{\ell+1} )}\big) g^{-1} V_{\ell+1/2}  \right\} , \\
V_{\ell + 1} &= V_{\ell+1/2} - (h/2) e^{\eta( t_{\ell+1} )} 
\Tr\left( \partial_X f(X_{\ell+1})  X_{\ell+1} T_a  \right) T_a .
\end{align}
\end{subequations}
This is a \emph{dissipative} version of leapfrog for Lie groups.
Finally, using the functions introduced in Eqs.~\eqref{alpha} and \eqref{beta}, 
and further rescaling
the velocity/momentum as
\be
e^{-\eta(t)}V(t) \to  V(t) ,
\ee
in analogy with \eqref{momchange},
we can  rewrite the method \eqref{lie_leap1} as
\begin{subequations} \label{lie_leap2}
\begin{align}
V_{\ell + 1/2} &= \alpha_{\ell+1/2} V_{\ell} - (h \alpha_{\ell+1/2}/2)  
\Tr\left( \partial_X f(X_\ell)  X_\ell  T_a  \right) T_a , \\
X_{\ell+1} &= X_\ell \exp\left( h  \beta_{\ell+1} g^{-1} V_{\ell+1/2}  \right) , \\
V_{\ell + 1} &= \alpha_{\ell+1} V_{\ell+1/2} - (h/2)  
\Tr\left( \partial_X f(X_{\ell+1})  X_{\ell+1}  T_a  \right) T_a .
\end{align}
\end{subequations}
The same previous comment regarding the mass $m=h$ applies
for solving optimization problems.
 Note also that we use  formula \eqref{trace_form} in place of the 
trace, so computations can be done in
global coordinates, i.e., we do not need to parametrize the group in terms of the generators, which is quite convenient in practice.
We thus state Algorithm~\ref{dissleaplie}.

\begin{algorithm}[t]
\caption{\label{dissleaplie}
\textsc{DissLeapfrogLie} is a presymplectic integrator for 
 solving optimization problems over Lie groups,
 $\min_{X \in \mathcal{G}} f(X)$.
 There is a sequence $\{\alpha_\ell \}$, where 
$\alpha_{\ell} \in (0,1)$ (momentum factor), and step size $h > 0$.
The parameter $\beta$ is fixed by Eq.~\eqref{beta}, and $g \succ 0$ is an
arbitrary constant matrix (preconditioner).}
\begin{algorithmic}[1]
\For{$\ell=0,1,\dotsc$}
\State $V_{\ell + 1/2} \leftarrow \alpha_{\ell+1/2} V_{\ell} - (h \alpha/2)  
\Tr\left[ \partial_X f(X_\ell)  X_\ell  T_a  \right] T_a $
\State $X_{\ell+1} \leftarrow X_\ell \exp\left(   \beta_{\ell+1} g^{-1} V_{\ell+1/2}  \right)$
\State $V_{\ell + 1} \leftarrow \alpha_{\ell+1} V_{\ell+1/2} - (h/2)  
\Tr\left[ \partial_X f(X_{\ell+1})  X_{\ell+1}  T_a  \right] T_a $
\EndFor
\end{algorithmic}
\end{algorithm}

Although $\alpha$ and $\beta$ are associated  to a generic damping
function $\eta(t)$, one can use the constant case
$\eta(t) = \gamma t$ as in Eq.~\eqref{alphabetaconst}.
Recall also that we can use formula \eqref{trace_form} to compute the trace,
without generators. Moreover,
 ``$\exp$'' refers to the matrix exponential, which defines a map from the Lie algebra to the Lie group. This exponential can  be replaced by any structure-preserving (symplectic) approximation such as a Cayley transform, 
\be
\exp( h Y) = \left(I - \tfrac{h}{2} Y\right)^{-1}\left(I + \tfrac{h}{2} Y\right) + \bigO(h^3).
\ee

Importantly, Algorithm~\ref{dissleaplie} can also be applied to naturally reductive \emph{homogeneous spaces}, which include:
\begin{itemize}
\item Stiefel manifolds. 
\item Grassmannian manifolds.
\item  The space of positive definite matrices 
(and their complex analogues). 
\item Projective spaces.
\item Affine spaces.
\end{itemize}

This is because we can view Hamiltonian systems on homogeneous spaces  as reduced mechanics obtained by Hamiltonian systems with symmetries on associated Lie groups.
Concretely, we simply need to restrict the momentum to a vector space complementary to the Lie algebra of the isotropy group, as discussed in the context of sampling \cite{Barp:2019}.

\section{Numerical Details for SSK}
\label{sec:numerics}

The problem \eqref{ssk_prob} is over a hypersphere, with the constraint
in the form 
\be\label{constr_sphere}
\gv(\bm{\sigma}) = \| \bm{\sigma}  \|^2 - n = 0.
\ee
Note that $\qq \equiv \bm{\sigma}$ in our notation.
Thus, the Jacobian matrix of constraints is simply 
$\Jac(\bm{\sigma})^T = 2 \bm{\sigma}$, and
the gradient of the objective function is $\nabla \mathcal{H}(\bm{\sigma}) = 
- M \bm{\sigma} - \rho \bm{g}$.
In fact, the constraint \eqref{constr_sphere} can be solved in closed form.
Consider Algorithm~\ref{dissgeorattle} and denote
\be
\qq_{\ell+1} = \bm{a} + \lambda \bm{b} ,
\ee
where $\bm{a} \equiv \qq_\ell + \beta g^{-1} \pp_{\ell}$,
$\bm{b} \equiv (\beta h \alpha / 2) g^{-1} \Jac(\qq_\ell)$,
and $\lambda \equiv \lambda_\ell$.
Note that in this case the Lagrange multiplier is a  scalar.
Solving \eqref{constr_sphere} for $\lambda$, i.e.,
computing line 7 in Algorithm~\ref{dissgeorattle}, requires
solving
\be
\| \bm{a} - \lambda \bm{b} \|^2 - n = 
\| \bm{b}\|^2 \lambda^2 - 2 \bm{a} \cdot \bm{b} \lambda + (\| \bm{a} \|^2  - n) = 0   ,
\ee
whose exact solution is (we should take the smallest root)
\be
\lambda_\ell = \dfrac{ \bm{a} \cdot \bm{b} - \sqrt{\Delta}}{ \| \bm{b}\|^2}, \qquad \Delta \equiv  (\bm{a} \cdot \bm{b})^2 -  \| \bm{b} \|^2 
\big( \| \bm{a}\|^2 - n \big).
\ee
We use this exact formula when solving problem 
\eqref{ssk_prob} with Algorithm~\ref{dissgeorattle}.\footnote{%
As a sanity check, we  compared this exact procedure with a numerical solution
of the constraint \eqref{constr_sphere} via Newton-Raphson method. 
The results were in complete agreement.}

In the examples of Fig.~\ref{fig:ssk} (Sec.~\ref{sec:numerical}) we
initialize  at
$\qq_0 = \bm{\sigma}_0 = \bm{1}$, which is on the $n$-sphere.
The momentum is always initialized at $\pp_0 = \bm{0}$.
Those examples show a single run of Algorithm~\ref{dissgeorattle}
in comparison with  gradient flow \cite{Sra:2016} (Riemannian gradient
descent). However, we verified similar results for several
instances of the problem as well. Note that the problem is of
dimension $n=1000$, which is significantly high dimensional and this
example provides a real  benchmark.

\begin{figure}[t]
\centering
\includegraphics[scale=0.7]{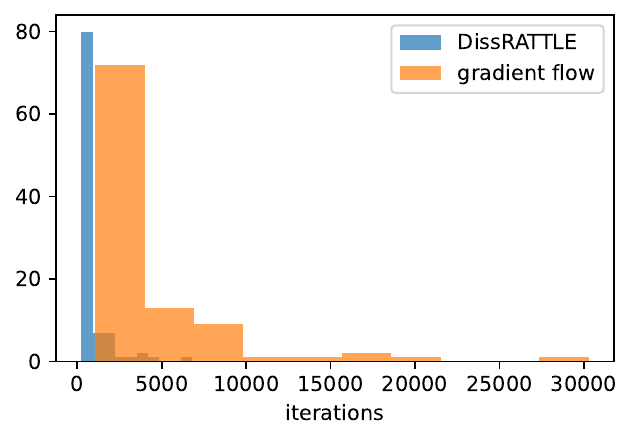}~~~~~
\includegraphics[scale=0.7,trim={0 -10 0 5}]{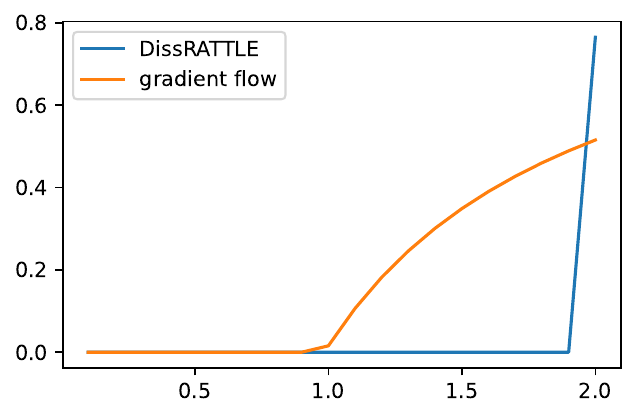}
\put(-110,-1){$C$}
\put(-225, 45){\rotatebox{90}{$|\mathcal{H}_\infty-\mathcal{H}_\star|/|\mathcal{H}_\star|$}}
\caption{\label{fig:hist_iter}
\emph{Left:} 100 Monte Carlo runs with random initializations on a problem
of dimension $n=500$.
We report histograms of the number of iterations such that  both
algorithms achieved a relative error $\sim 10^{-14}$ in the objective
function value. Algorithm~\ref{dissgeorattle} is much
faster than gradient flow (Riemannian gradient descent \cite{Sra:2016}).
Both methods used the same step size, favoring gradient flow.
\emph{Right:}
\label{fig:step_thre}
Plot of the relative error against $C$, where
$h  = C / \lambda_{\text{max}}(M)$.
Note that Algorithm~\ref{dissgeorattle} is almost twice
times more stable than gradient flow, even though it is
``accelerated.''
}
\end{figure}

To provide error quantification of our method, we 
consider 100 Monte Carlo runs of Algorithm~\ref{dissgeorattle} and
gradient flow, both with step size
$h = 0.5/\lambda_{\text{max}}(M)$. We fix the momentum factor as $\alpha=0.9$.
We consider random initializations, where 
we choose one component of $\bm{\sigma}_0$ at random, say the $i$th component,
and set it to $\sigma^i_0 = \sqrt{n}$, while the remaining components
are $\sigma^j_0 = 0$ for $j \ne i$.
We run both methods up to a tolerance
error of $10^{-10}$ (which provides a relative error with respect
to the minimum of the objective
function of $\sim 10^{-14}$). We consider problem \eqref{HSSK} with
$\rho=0$ (no external field) in $n=500$ dimensions, which
is large. We show in Fig.~\ref{fig:hist_iter} (left)
a histogram of the number of iterations.
All methods, and in all instances of the problem, converged successfully
to the required accuracy.
Note, however, that Algorithm~\ref{dissgeorattle} is consistently 
faster than gradient flow by orders of magnitude.

\begin{figure}[t]
\centering
\includegraphics[scale=.8,trim={0 0 0 0}]{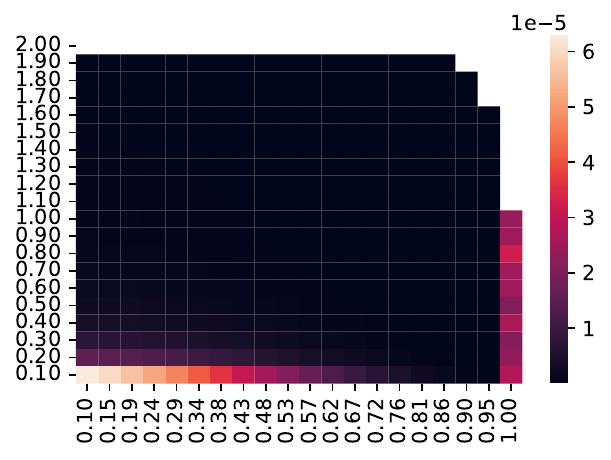}
\put(-120,-5){$\alpha$}
\put(-250, 90){$C$}
\caption{\label{fig:heat}
We vary the parameter $\alpha \in (0,1)$ and the step size 
$h = C / \lambda_{\text{max}}(M)$,
with 
$C \in (0,2)$, into Algorithm~\ref{dissgeorattle} for an
instance of problem \eqref{ssk_prob}.
The heatmap is the value of the relative error
$|\mathcal{H}_\infty-\mathcal{H}_\star|/|\mathcal{H}_\star|$.
The white area corresponds to cases where the
algorithm did not converge.
}
\end{figure}

Next, for a fixed instance of the problem in $n=200$ dimensions, as described in the
previous paragraph, we fix $\alpha=0.9$ and consider a step size
$h = C / \lambda_{\text{max}}(M)$, where we vary 
$C \in [0.1, 2]$. 
We require all algorithms to achieve a tolerance error
$\sim 10^{-7}$ in the relative error of the objective function
value with respect to the optimum.
We show in Fig.~\ref{fig:step_thre} (right) a plot of this relative
error against $C$.
Note how Algorithm~\ref{dissgeorattle} is considerable
more stable than gradient flow (Riemannian gradient descent).
We repeated this experiment for several instances
of the problem, obtaining identical results as the one
displayed.

Finally, we now vary both parameters
$\alpha$ and the step size $h$ into 
Algorithm~\ref{dissgeorattle} for an
instance of problem \eqref{ssk_prob} (with $\rho=0$, no external
field) in $n=100$ dimensions.
This allows us to find the stability region of the
method in the parameter space.
The results are shown in the heatmap of Fig.~\ref{fig:heat}.
For the majority of parameter choices the algorithm
converged successfully.
Only for very large values of $C \simeq 2$ the method diverged
(white area), or when the method was very weakly damped,
$\alpha \simeq 1$.
Interestingly, the method converged  even for large
values of $\alpha$. The entire dark region corresponds to
values of the relative error of $\sim 10^{-6}$, according to our
chosen tolerance error.
We repeated this experiment for few other instances
of the problem, with no noticeable change in the results.
Therefore, Algorithm~\ref{dissgeorattle} proved to be quite
stable.

\subsection{SSK over $SO(n)$}

\begin{figure}[t]
\centering
~~~\includegraphics[scale=0.55]{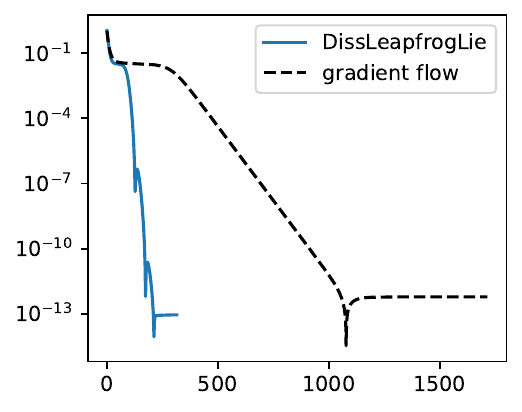}~~~~~
\includegraphics[scale=0.55]{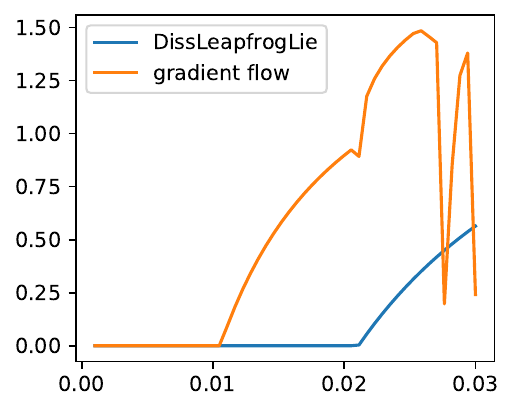}~~
\includegraphics[scale=0.55]{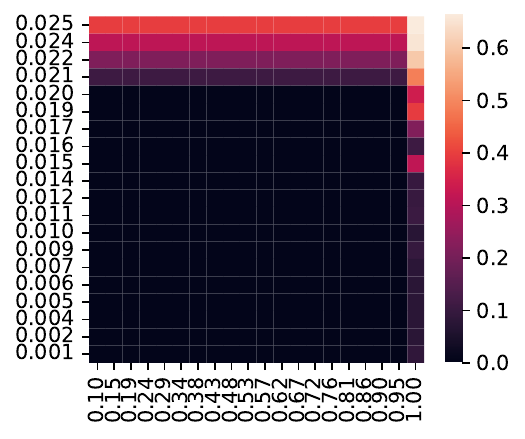}
\put(-450,25){\rotatebox{90}{$|\mathcal{H}_\ell-\mathcal{H}_\star|/|\mathcal{H}_\star|$}}
\put(-390, -5){iterations ($\ell$)}
\put(-295,25){\rotatebox{90}{$|\mathcal{H}_\infty-\mathcal{H}_\star|/|\mathcal{H}_\star|$}}
\put(-215,-5){$C$}
\put(-148,60){$C$}
\put(-70,-5){$\alpha$}
\caption{\label{fig:ssk_lie} 
Solving problem \eqref{ssk_lie} with Algorithm~\ref{dissleaplie} and
Riemannian gradient descent (gradient flow) \cite{Sra:2016}.
\emph{Left:} We choose
$n=500$, $\alpha=0.9$, and step size 
$h=0.001/\lambda_{\text{max}}(M)$ so gradient flow could converge.
\emph{Middle:} We vary the step size $h = C / \lambda_{\text{max}}(M)$.
Note how Algorithm~\ref{dissleaplie} is twice more stable.
\emph{Right:} Heat map of the relative error in the objective function
by varying $\alpha$ and $C$ for Algorithm~\ref{dissleaplie}. The method
was able to converge for a large range of values in the parameter space.
}
\end{figure}
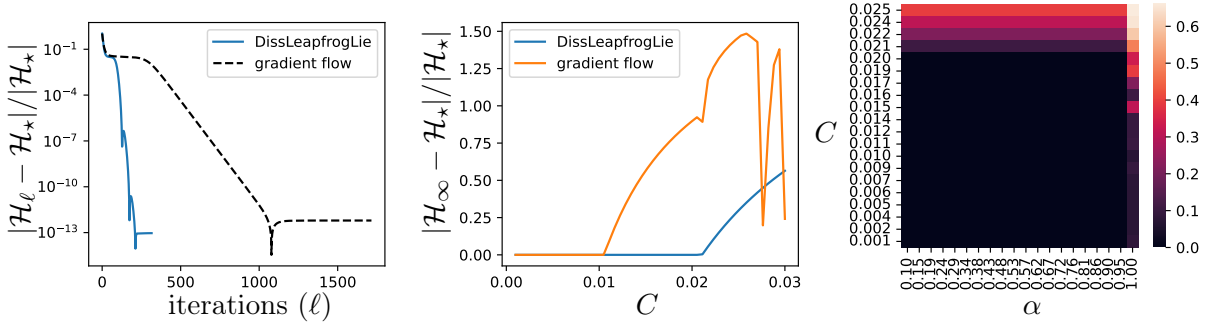

It is possible to redefine problem \eqref{HSSK}/\eqref{ssk_prob}
over a homogenous space.
This is done as
\be\label{ssk_ham_lie}
\mathcal{H}(X) = -\dfrac{1}{2} \sum_{i,j=1}^n X_{i1} M^{ij} X_{j1}
- \rho \sum_{i=1}^n g^i X_{i1}
\ee
where $X \in SO(n)$, the special orthogonal group.
This is because the oriented sphere can be written as
$\mathcal{S}_{n-1} \equiv SO(n)/SO(n-1)$.
We can thus find the ground state of an SSK model by  an optimization problem
over the $SO(n)$ group,
\be\label{ssk_lie}
\min_{X \in SO(n)} \mathcal{H}(X).
\ee
We  apply Algorithm~\ref{dissleaplie} to this problem, with $\rho=0$ (no external field) and damping $\eta(t) = \gamma t$; see Eq.~\eqref{alphabetaconst}. Similar
results are obtained for an external field, $\rho\ne 0$. We compare this method with Riemannian gradient descent \cite{Sra:2016} 
(``gradient flow''), when the geodesic flow is computed over the Lie group.
In Fig.~\ref{fig:ssk_lie} (left) we show an instance of this problem.
We can see that Algorithm~\ref{dissleaplie} is significantly faster, besides
 being more stable (middle plot).
We  show the stability of the method on parameter space (right plot); 
the heatmap has values $|\mathcal{H}_\infty - \mathcal{H}_\star|/|\mathcal{H}_\star|$.

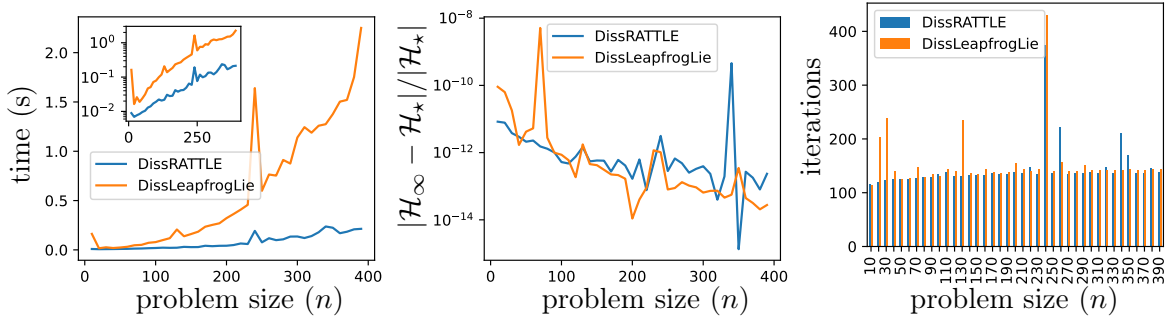
\begin{figure}[t]
\centering
~~\includegraphics[scale=0.55]{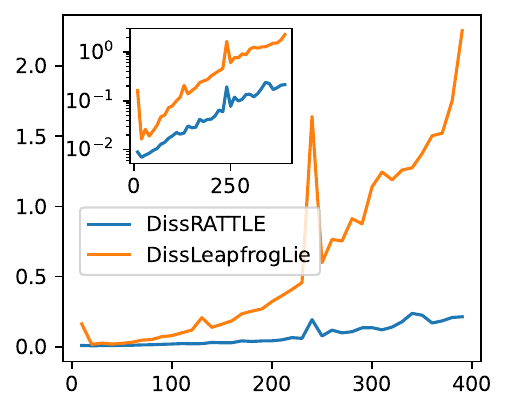}~~
\includegraphics[scale=0.55]{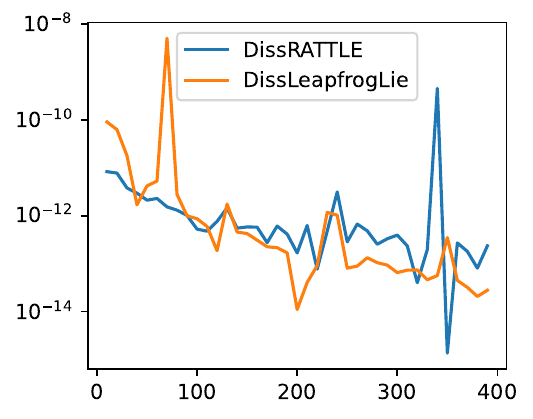}~
\includegraphics[scale=0.55]{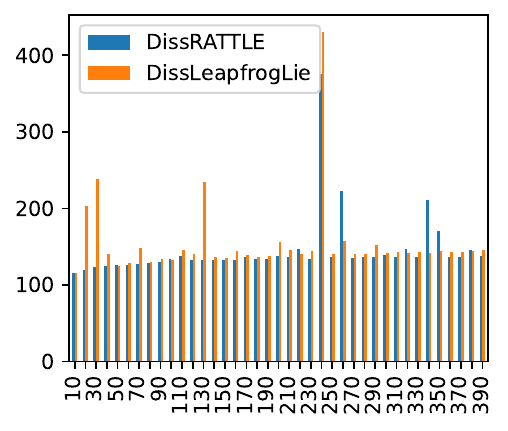}
\put(-438,42){\rotatebox{90}{time (s)}}
\put(-390, -5){problem size ($n$)}
\put(-292,25){\rotatebox{90}{$|\mathcal{H}_\infty-\mathcal{H}_\star|/|\mathcal{H}_\star|$}}
\put(-238,-5){problem size ($n$)}
\put(-141,45){\rotatebox{90}{iterations}}
\put(-100,-5){problem size ($n$)}
\caption{\label{fig:constr_vs_lie} 
Comparing Algorithms~\ref{dissgeorattle}
and \ref{dissleaplie} when solving the SSK problem.
\emph{Left:} Wall-clock time vs. $n$. 
The inset shows the $y$-axis in log scale.
Note how Algorithm~\ref{dissgeorattle}
is more scalable and also stable. Algorithm~\ref{dissleaplie} diverged
for higher $n$ (with this choice of step size), contrary to Algorithm~\ref{dissgeorattle} which could
operate with even larger step sizes.
\emph{Middle:} We show the relative error of the objective
function to confirm that both methods achieved the same accuracy in the solution.
\emph{Right:} We also show the number of iterations, which is also similar
for both methods with this choice of parameters.
}
\end{figure}

In high dimensions, applying the constrained method 
in Algorithm~\ref{dissgeorattle} is more efficient than the matrix approach
of Algorithm~\ref{dissleaplie}.
To illustrate this, in Fig.~\ref{fig:constr_vs_lie} 
we compare both methods when applied to the same SSK problem
and with the same initial condition. We vary the size $n$ of the problem. 
For both methods we set $\alpha=0.9$.
The step sizes were chosen so that
both methods achieved approximately the same tolerance error during the
same number of iterations.
For Algorithm~\ref{dissgeorattle} we set
$h=1/\lambda_{\text{max}}(M)$ and for Algorithm~\ref{dissleaplie} we set
$h=0.005/\lambda_{\text{max}}(M)$. However, Algorithm~\ref{dissleaplie} became
unstable for larger problem sizes (with this step size choice),
while Algorithm~\ref{dissgeorattle} did not and in fact 
could operate with even larger step sizes.
Overall, we believe the approach through constrained optimization
is superior and more scalable whenever the problem can be vectorized efficiently.

\section{Orthogonal Procrustes Problem}

We consider the problem
\be
\label{whaba_prob}
\min_{X \in SO(n)} \| M - X \|_F^2 ,
\ee
where $M$ is a given matrix. Note that the constraints are
$X^T X  = X X^T = I$ and $\det X = +1$, so this  is a \emph{nonconvex} problem. 
This is a generalization of
the famous orthogonal Procrustes problem, which relaxes the condition
$\det X = +1$;
it has several applications in statistics, multidimensional scaling,
and natural language processing \cite{Gower:2004}. 
Problem \eqref{whaba_prob} is also known as Wahba's problem
\cite{Wahba:1965}.
There is a closed form solution to this problem, which however
involves computing an SVD: $X_\star = U D V^T$, where $M = U \Sigma V^T$, and
$D_{ij} = 0$ for $i\ne j$, $D_{ii} = 1$ for $i=1,\dotsc,n-1$, and
$D_{nn} = \det(U V^T)$.

\begin{figure}[t]
\centering
~~\includegraphics[scale=0.58]{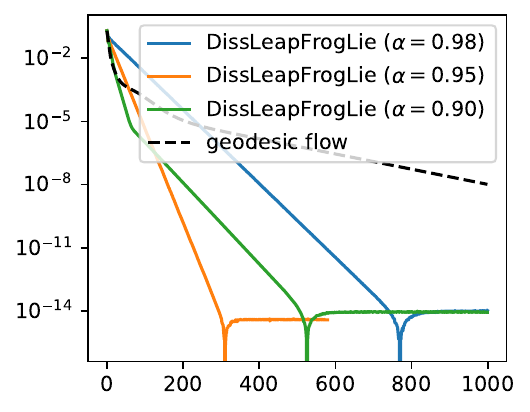}%
\includegraphics[scale=0.58]{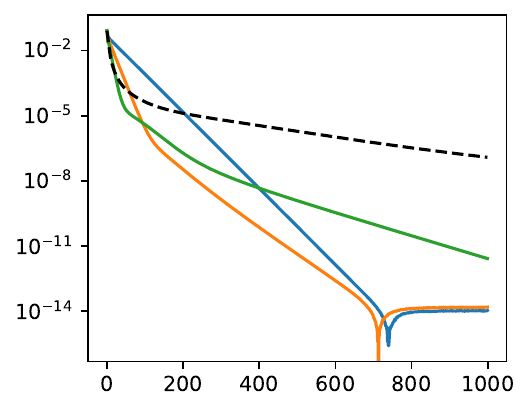}%
\includegraphics[scale=0.58]{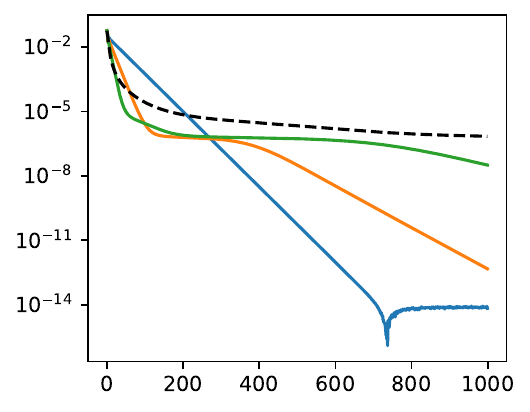}
\put(-450,30){\rotatebox{90}{$|f_\ell - f_\star|/|f_\star|$}}
\put(-390, -5){iteration ($\ell$)}
\put(-240, -5){iteration ($\ell$)}
\put(-90, -5){iteration ($\ell$)}
\caption{\label{fig:whaba} 
Solving problem \eqref{whaba_prob} with Algorithm~\ref{dissleaplie}
and Riemannian gradient descent (gradient flow).
In all cases we sample $M_{ij} \sim \mathcal{N}(0,1)$.
\emph{Left:} Problem with dimension $n=100$.
\emph{Middle:} $n=500$.
\emph{Right:} $n=1000$.
If we carefully tune $\alpha$ and $h$ in Algorithm~\ref{dissleaplie}
the convergence is even faster than shown in these plots.
Note how the convergence of gradient flow is quite slow, specially
for large $n$.
}
\end{figure}

\begin{figure}[t]
\centering
\includegraphics[scale=0.7]{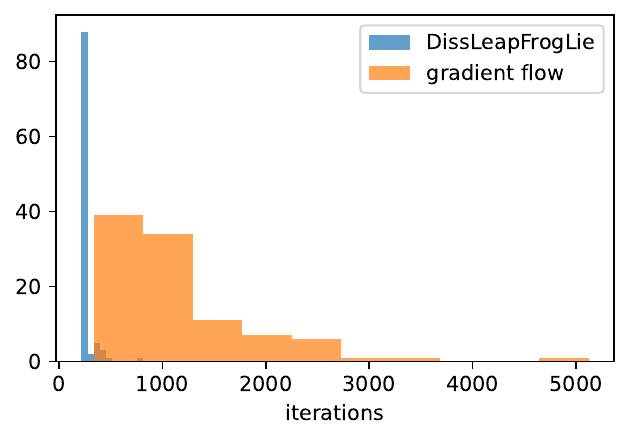}
\includegraphics[scale=0.7]{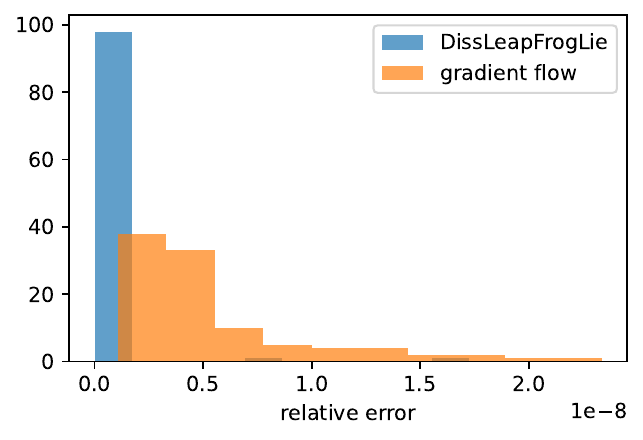}
\caption{\label{fig:whaba_hist} 
We consider 100 Monte Carlo runs on problem \eqref{whaba_prob} in
$n=100$ dimensions, with
Algorithm~\ref{dissleaplie} ($\alpha=0.95$) and Riemannian gradient
descent; both with step size $h = \tfrac{1}{2} \lambda_{\text{max}}(M)$.
\emph{Left:} Histogram of the number of iterations for convergence
up to a tolerance error $\approx 10^{-8}$.
\emph{Right:} Histogram of the relative error of the final iteration.
Note how Algorithm~\ref{dissleaplie} converges in a much smaller number
of iteration, and to a more accurate solution.
}
\end{figure}

\begin{figure}[t]
\centering
\includegraphics[scale=0.68]{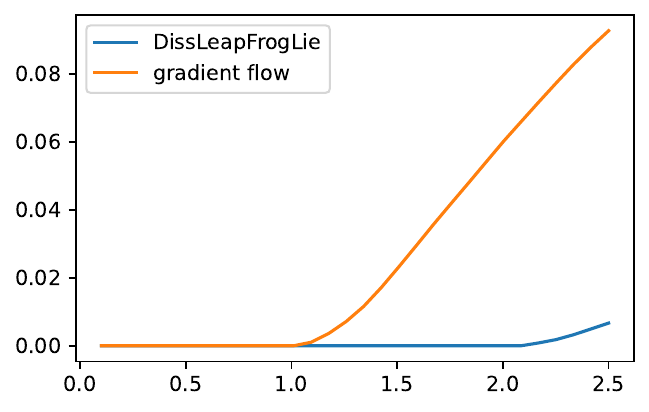}~~
\includegraphics[scale=0.68]{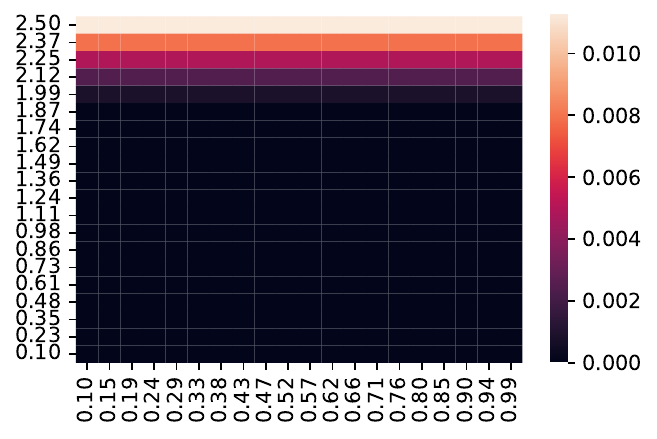}
\put(-330,-5){$C$}
\put(-445,35){\rotatebox{90}{$|f_\ell - f_\star|/|f_\star|$}}
\put(-225,80){$C$}
\put(-135,-5){$\alpha$}
\caption{\label{fig:params} 
Problem \eqref{whaba_prob} in $n=50$ dimensions.
\emph{Left:} We set $\alpha=0.95$ into Algorithm \ref{dissleaplie}
and vary $C \in (0, 2.5)$ for step size $h = C \lambda_{\text{max}}(M)$.
Note how Algorithm~\ref{dissleaplie} is much more stable than
gradient flow.
\emph{Right:} For another instance of the problem,
we check the stability of Algorithm~\ref{dissleaplie} by varying
$C$ and $\alpha$; the heatmap shows values of 
the achieved relative error of the objective
function after convergence up to tolerance $10^{-8}$ (in the state variable) or 2000 iterations.
}
\end{figure}

Problem \eqref{whaba_prob} is well-suited 
to Algorithm~\ref{dissleaplie}, whose implementation becomes extremely simple.
We thus compare this method with Riemannian gradient descent (gradient flow),
for a few values of $n$. We choose three different values
of $\alpha$ to illustrate the role of the momentum factor.
The results are shown in Fig.~\ref{fig:whaba} for a single run
of these algorithms, with initialization $X_0 = I$.
The step size of all methods is chosen as
$h = \tfrac{1}{2} \lambda_{\text{max}}(M)$, which allowed gradient
flow to converge; Algorithm~\ref{dissleaplie} allows much larger
step sizes however, which makes its convergence faster than shown
in these plots.
Note how gradient flow gets stalled, specially with increasing dimensionality.
Next, In Fig.~\ref{fig:whaba_hist}, we consider 100 Monte Carlo runs
for $n=100$ and show histograms of the number of iterations and
the relative error in the objective function, running both methods
until achieving a tolerance error $\approx 10^{-8}$ in the solution.
In Fig.~\ref{fig:params} we illustrate the improved 
stability of Algorithm \ref{dissleaplie}
compared to Riemannian gradient descent.

\bibliography{biblio.bib}

\end{document}